\documentclass{article}
\usepackage[utf8]{inputenc}
\usepackage{enumitem}
\usepackage{hyperref}
\usepackage{authblk}
\usepackage{xcolor}
\usepackage{geometry}
\usepackage{amssymb,amsmath}
\usepackage{graphicx}
\usepackage{bm}
\usepackage{biblatex}
\usepackage{pgf,tikz}
\usetikzlibrary{positioning}
\usepackage{amsthm}
\usepackage[ruled,vlined,linesnumbered]{algorithm2e}

\usepackage{booktabs}
\usepackage{longtable}
\usepackage{array}
\usepackage{multirow}
\usepackage{wrapfig}
\usepackage{float}
\usepackage{colortbl}
\usepackage{pdflscape}
\usepackage{tabu}
\usepackage{threeparttable}
\usepackage{threeparttablex}
\usepackage{ifxetex,ifluatex}
\usepackage[section]{placeins}
\usepackage{adjustbox}

\newtheorem{theorem}{Theorem}
\newtheorem{corollary}{Corollary}

\allowdisplaybreaks
\bibliography{ref.bib}

\title{Targeted Learning Estimation of Sampling Variance for Improved Inference}
\date{\today}
\author[1]{Yunwen Ji}
\author[1]{Mark van der Laan}
\author[1]{Alan Hubbard}
\affil[1]{Division of Biostatistics, University of California, Berkeley}

\begin{document}

\maketitle

\begin{abstract}
    For robust statistical inference it is crucial to obtain a good estimator of the variance of the proposed estimator of the statistical estimand. A commonly used estimator of the variance for an asymptotically linear estimator is the sample variance of the estimated influence function. This estimator has been shown to be anti-conservative in limited samples or in the presence of near-positivity violations, leading to elevated Type-I error rates and poor coverage. In this paper, capitalizing on earlier attempts at targeted variance estimators, we propose a one-step targeted variance estimator for the causal risk ratio (CRR) in scenarios involving treatment, outcome, and baseline covariates. While our primary focus is on the variance of log(CRR), our methodology can be extended to other causal effect parameters. Specifically, we focus on the variance of the IF for the log relative risk (log(CRR)) estimator, which requires deriving the efficient influence function for the variance of the IF as the basis for constructing the estimator. Several methods are available to develop efficient estimators of asymptotically linear parameters. In this paper, we concentrate on the so-called one-step targeted maximum likelihood estimator, which is a substitution estimator that utilizes a one-dimensional universal least favorable parametric submodel when updating the distribution. We conduct simulations with different effect sizes, sample sizes and levels of positivity to compare the estimator with existing methods in terms of coverage and Type-I error. Simulation results demonstrate that, especially with small samples and near‑positivity violations, the proposed variance estimator offers improved performance, achieving coverage closer to the nominal level of 0.95 and a lower Type-I error rate. 
   
\end{abstract}

\section{Introduction}\label{sec:introduction}

Numerous estimators are available for the estimation of causal effects. Among these, semiparametric efficient and doubly robust estimation methods stand out for their ability to yield reliable estimates particularly in the context of high-dimensional data \cite{bickel1998efficient, LaanTLCI2011, tsiatis2006semiparametric}. Estimators such as the one-step estimator and the Targeted Maximum Likelihood Estimator (TMLE) provide enhanced flexibility and robustness, as they do not require a fully specified parametric form of the data model and make use of the efficient influence function (EIF) \cite{bickel1998efficient, vanderLaanRubin2006, vanderLaanGruber2011}. These estimators have demonstrated their power and reliability across various real-world applications, particularly in public health and clinical research, where they effectively handle the complexities inherent in observational data \cite{li_evaluating_2022}. 

Regular and asymptotic linear estimators (RAL) imply that, in large enough samples, the estimator has an approximately normal sampling distribution and converges to the true parameter value. Therefore, the variance of an RAL estimator can be consistently estimated with the sample variance of the estimated influence function of the RAL estimator. However, under violations of positivity, such variance estimators can be anti-conservative, resulting in uncontrolled Type I errors and biased confidence interval coverage \cite{vanderLaanGruber2011, petersen_diagnosing_2012}. Petersen et al. \cite{petersen_diagnosing_2012} emphasized that data sparsity can exacerbate bias and variance in causal effect estimators and proposed using a parametric bootstrap to assess bias arising from positivity violations and finite samples. Tran et al. \cite{tran_robust_2018} introduced a semi-targeted estimation of the variance of the efficient influence function (EIF) of treatment-specific means, constructing and estimating these means by semi-targeted learning rather than fully targeting the parameter of interest. In addition, Coyle et al. \cite{van2018targeted} presented an iterative TMLE for the variance of the EIF for the treatment-specific mean by making multiple sequential updates to the initial estimate until convergence.  Although this method directly targets the variance, it still gives anti-conservative results in simulations of small sample sizes due to the iterative fitting of data in the targeting step \cite{van2018targeted}. 

In this paper, we present a one-step TMLE based on a one-dimensional universally favorable model to estimate the variance of the EIF of the log causal risk ratio (log(CRR)), while the idea can be applied to other causal parameters. The proposed estimator utilizes a one-dimensional universally least favorable submodel (ULFM) as described by van der Laan et al \cite{van_der_laan_one-step_2016}. Iterative TMLE is a two-stage estimator that involves an initial estimate of the components of the data-generating distribution that require fitting/smoothing (e.g., the outcome regression) followed by a targeted step that locally maximizes the change in the target parameter. However, this choice of submodel is tailored to be optimal locally around the initial estimator, so that its ability to provide maximal per unit change in the likelihood of the fit relies on the initial estimator being close enough to the true probability distribution. In contrast, one-step TMLE is able to achieve the target in one-step by using a univerally least favorable model (ULFM), with minimal additional fitting of the data. This method avoids the potential overfitting that can occur with iterative TMLE, particularly in small sample sizes. Previous studies have demonstrated the robustness and efficiency of one-step TMLE in longitudinal and survival analysis settings, highlighting its ability to improve finite-sample robustness  \cite{cai2020one,rytgaard2024one}.

In this paper, we compare the performance of four variance estimators: (1) the empirical variance of the estimated EIF, (2) a simple substitution estimator, (3) an iterative TMLE, and (4) a one-step TMLE. Our results show that confidence intervals based on the one-step TMLE provide consistent nominal coverage across various data-generating distribution (DGD) scenarios, particularly outperforming the other estimators in smaller samples and under DGDs with positivity violations. We further illustrate our approach through a real-data based simulation, modified to ensure certain degrees of positivity violations, where the one-step TMLE provides more conservative and reliable variance estimates compared to the empirical variance estimator.

\subsection{Organization of the paper}

In Section \ref{sec:target}, we introduce our target parameter, the variance of the efficient influence function of log(CRR). Section \ref{sec:Estimation} presents existing estimators and proposes a new one-step TMLE for variance, using a one-dimensional universal least favorable parametric submodel to target the parameter. In Section \ref{sec:simulation}, we describe the settings for our simulations. We then compare the performance of four estimators under varying effect sizes, sample sizes, and levels of positivity. We apply our methodology to a real-world example in Section \ref{sec:data}. Finally, in Section \ref{sec:Discussion}, we discuss the limitations of this study and propose directions for future research.
\section{Definition of data and statistical estimand}\label{sec:target}
\subsection{Data and Notation}
We observe $n$ independent and identically distributed copies of $O$ with a data structure $O = (W, A, Y) \sim{P_0} \in \mathcal{M}$, where $W \in  \mathbb{R}^p$ denotes $p$-dimensional covariates, $A \in \{0, 1\}$ the binary treatment variable,  $Y \in \{0, 1\}$ the binary outcome. The statistical model for the probability distribution $P_0$  is denoted by $\mathcal{M}$. 

 We use the following structural causal model (SCM) \cite{pearl2000causality}:

$$
\begin{aligned}
 W&=f_W\left(U_W\right) \\ 
A&=f_A\left(W, U_A\right) \\ 
Y&=f_Y\left(W, A, U_Y\right),
\end{aligned}
$$

where $U$ denotes exogenous errors and $f$ deterministic functions.
 
We denote the treatment-specific counterfactual mean as $\mathbb{E}(Y^a)$, where $Y^a$ is the counterfactual result under treatment $A=a$, with $a$ being either 0 (untreated) or 1 (treated). In this paper, we focus on the binary outcome case. The causal parameter, log of the causal risk ratio (log(CRR)) on the full data is:
 $$
 \Psi^{F}=\text{log} \left(\frac{\mathbb{E}(Y^1)}{\mathbb{E}(Y^0)}\right) = \text{log}(\mathbb{E}Y^1) - \text{log}(\mathbb{E}Y^0).
 $$

If we make the standard assumptions in causal inference, such as no unmeasured confounding $(\{Y^0, Y^1\} \perp A | W)$, positivity $(0 < P(A = 1|W) < 1)$ and consistency, then the full-data causal parameter $\Psi^F$ can be identified by the following statistical parameter $\Psi:\mathcal{M} \rightarrow \mathbb{R}$ defined by:  
\begin{equation} \label{eq:causal_param}
\begin{aligned}
     \Psi^{F} = \Psi(P_0)&=\text{log } \mathbb{E}_{P_0}(\mathbb{E}_{P_0}(Y \mid W, A=1)) - \text{log }\mathbb{E}_{P_0}(\mathbb{E}_{P_0}(Y \mid W, A=0)).
\end{aligned}
\end{equation}

Note that the causal model and the identification are not relevant for the development of the following variance estimator. Our target parameter is a statistical parameter, which allows us to construct a variance estimator for an efficient estimator of that statistical estimand in Equation \ref{eq:causal_param} . Under causal assumptions, $\Psi(P_0)$ equals $\text{log} \frac{EY^1}{EY^0}$, but this causal interpretation is not necessary for either the construction of TMLE or the evaluation of its coverage properties. Thus, no further causal interpretation will be discussed from this point forward.

\subsection{Target Parameter}
In this section, we focus on the target parameter of interest: the variance of the efficient influence function (EIF) of a casual parameter, which is the log of the causal relative risk (log(CRR)). Although the choice of the association parameter is somewhat arbitrary, this framework we present can be applied equivalently to other estimands of interest.

\subsubsection{Variance of EIF of log(CRR)}
We define $\Psi(P) = \text{log } \psi_1(P)-\text{log } \psi_0(P)$, where $\psi_a(P)=\mathbb{E}Y^a$. Let $\bar{Q}_P(A, W)  \equiv E_P(Y \mid W, A), g_P(A \mid W)  \equiv P(A \mid W)$. And we use the notation $\bar{Q}_0(A, W) \equiv \bar{Q}_{P_0}(A, W)$ and $ g_0(A\mid W)\equiv g_{P_0}(A\mid W)$ for the distribution $P_0$. 

Given that the efficient influence function of $\psi_a$ at $P$ is
$$
D_{\psi_a, P}^*(O) = \frac{\mathbb{I}(A=a)}{g_P(a \mid W)}\left(Y-\bar{Q}_P(A, W)\right)+\bar{Q}_P(a, W)-\psi_a(P)
$$ 

and $\Psi(P) = \text{log } \psi_1(P)-\text{log } \psi_0(P)$, we can apply the delta method \cite{LaanTLCI2011} to derive the efficient influence function of  $\Psi: \mathcal{M} \rightarrow \mathbb{R}$ at $P$ as follows:
\begin{equation}\label{eq:IC_logRR}
    D_{\Psi, P}^*(O) = \left(\frac{\mathbb{I}(A=1)}{\psi_1(P) g_{P}(1 \mid W)}-\frac{\mathbb{I}(A=0)}{\psi_0(P) g_{P}(0 \mid W)}\right)\left(Y-\bar{Q}_{P}(A, W)\right)+\frac{\bar{Q}_{P}(1, W)}{\psi_1(P)}-\frac{\bar{Q}_{P}(0, W)}{\psi_0(P)}.
\end{equation}

Let $\Psi_0 = \Psi(P_0)$ denote the statistical parameter, and let $\Psi_n = \Psi(P_n)$ represent its asymptotically linear estimator based on the empirical distribution $P_n$. An estimator $\Psi(P_n)$ is asymptotically efficient if and only if it is asymptotically linear with an influence function (IF) equal to the efficient influence function (EIF) $D_{\Psi, P_0}^*(O)$:
\begin{equation}
    \Psi{(P_n)}-\Psi{(P_0)}\,=\,\frac{1}{n} \sum_{i=1}^{n} D_{\Psi, P_0}^*\left(O_{i}\right)+o_{P}\left(n^{-1 / 2}\right).
\end{equation}

Our estimand of interest is the {\em variance} of the EIF of log(CRR), which can be defined as a statistical parameter mapping $\Sigma^2: \mathcal{M} \rightarrow \mathbb{R}$:

\begin{equation}\label{eq:variance}
\begin{split}
    \Sigma^2(P) &= P\{D_{\Psi, P}^*(O)\}^2 \\
    & = \mathbb{E}_{W}\left(\frac{1}{\left(\psi_1(P)\right)^2} \frac{\left(\bar{Q}_P(1, W)\left(1-\bar{Q}_P(1, W)\right)\right)}{g_P(1 \mid W)}+\frac{1}{\left(\psi_0(P)\right)^2} \frac{\left(\bar{Q}_P(0, W)\left(1-\bar{Q}_P(0, W)\right)\right)}{g_P(0 \mid W)}\right.\\
    &\quad+\left.\left(\frac{\bar{Q}_P(1, W)}{\psi_1(P)}-\frac{\bar{Q}_P(0, W)}{\psi_0(P)}\right)^2\right).
\end{split}
\end{equation}

We define $\Sigma^2_0 = \Sigma^2(P_0)$, which is the true value of the target estimand of interest. The derivation of Equation \ref{eq:variance} can be found in Appendix \ref{appendix: aA}. 
\section{Estimation} \label{sec:Estimation}

We propose a one-step Targeted Maximum Likelihood Estimator (TMLE) for the variance of the EIF of the log causal relative risk (log(CRR)) as defined in Equation \ref{eq:variance}. 
To provide context, we begin by describing two commonly used variance estimators in Section \ref{sec:others}, the influence function-based variance estimator and the non-targeted substitution estimator.

Suppose we have $n$ observations $O_i=(W_i, A_i, Y_i)$ drawn independently from $P_0$. We use the subscript $n$ to indicate the estimator based on the empirical distribution, $P_n$ (e.g., $\bar{Q}_n(A, W), g_n(A|W)$). Superscripts indicate updates during the targeted learning process, where $(0)$ refers to initial estimates (E.g. $\bar{Q}^{(0)}_n(A, W), g^{(0)}_n(A|W)$) ,  $(*)$ the TMLE-updated estimates (E.g. $\bar{Q}^{(*)}_n(A, W), g^{(*)}_n(A|W)$) for the causal parameter $\Psi_0$, and (**) the TMLE-updated estimates for the target parameter of interest $\Sigma^2_0$.

\subsection{Review of existing estimators} \label{sec:others}

\subsubsection{Influence function-based estimator}
The empirical variance estimator based on the empirical influence function of $\Psi$ is defined as:
\begin{equation}
    \Sigma^{2, IC}_n  = \frac{1}{n} \sum_{i=1}^{n} \left(D_{\Psi, P^*_n}^{*}\left(O_{i}\right)\right)^{2},
\end{equation}
where $D_{\Psi, P^*_n}^{*}\left(O_{i}\right)$ is the efficient influence function of $\Psi$ defined in Equation \ref{eq:IC_logRR}, evaluated at the distribution $P^*_n$ for the observation $O_i$. $P^*_n$ is the distribution evaluated at the targeted estimates of $P_0$ with $\bar{Q}^{(*)}_n(A, W)$ and $g^{(*)}_n(A|W)$. \\


\subsubsection{Non-targeted (simple) substitution estimator}
\noindent By Equation \ref{eq:variance}, the non-targeted substitution (SS) estimator is defined as: 
\begin{equation}
\begin{split}
     \Sigma_{n}^{2, SS} & ={\frac{1}{(\psi_{1,\,n}^{(0)})^{2}}\frac{1}{n} \sum_{i=1}^{n} \frac{\bar{Q}_{n}^{(0)}\left(1,\,W\right)\left(1-\bar{Q}_{n}^{(0)}\left(1,\,W\right)\right)}{g_n^{(0)}(1\mid W)}+\frac{1}{(\psi_{0,\,n}^{(0)})} \frac{1}{n} \sum_{i=1}^{n} \frac{\bar{Q}_{n}^{(0)}\left(0,\,W\right)\left(1-\bar{Q}_{n}^{(0)}\left(0,\,W\right)\right)}{g_n^{(0)}(0|W)}}\\
     &\quad+{\frac{1}{n}\sum_{i=1}^{n}\left(\frac{\bar{Q}_{n}^{(0)}\left(1,\,W\right)}{\psi_{1,\,n}^{(0)}}-\frac{\bar{Q}_{n}^{(0)}\left(0,\,W\right)}{\psi_{0,\,n}^{(0)}}\right)^{2}}
\end{split}
\end{equation}
where  $\bar{Q}_{n}^{(0)}$, $g_{n}^{(0)}$, $\psi_{1,\,n}^{(0)}$ and $\psi_{0,\,n}^{(0)}$ are the initial estimates obtained using the SuperLearner, which will be discussed in detail in Section \ref{sec:initial_est}. This estimator provides a more sensitive measure of the variance than the empirical variance of the EIF, as its analytic form (Equation \ref{eq:variance}) explicitly integrates terms associated with rare observations, thereby avoiding the underestimation that can occur with the empirical method. However, unlike TMLE, it does not include a targeting step that reduces bias, making it less robust.

\subsection{Targeted maximum likelihood estimator} \label{tmle}
We introduce the TMLE approach for $\Sigma^2_0$ and its corresponding efficient influence function (EIF). TMLE is a plug-in estimator involving two key steps: initial estimation and targeting. The targeting step plays a crucial role in correcting plug-in bias and ensuring an asymptotically normal sampling distribution by refining the initial estimates $\bar{Q}_n^{(0)}$ and $g_n^{(0)}$.  \\

We propose the following TMLE $\Sigma_{n}^{2, TMLE}$ for $\Sigma^{2}_{0}$:

\begin{equation} \label{eq: tmle_est}
\begin{split}
     \Sigma_{n}^{2, TMLE} & ={\frac{1}{(\psi_{1,\,n}^{(**)})^{2}}\frac{1}{n} \sum_{i=1}^{n} \frac{\bar{Q}_{n}^{(**)}\left(1,\,W\right)\left(1-\bar{Q}_{n}^{(**)}\left(1,\,W\right)\right)}{g_n^{(**)}(1|W)}+\frac{1}{(\psi_{0,\,n}^{(**)})} \frac{1}{n} \sum_{i=1}^{n} \frac{\bar{Q}_{n}^{(**)}\left(0,\,W\right)\left(1-\bar{Q}_{n}^{(**)}\left(0,\,W\right)\right)}{g_n^*(0|W)}}\\
     &\quad+{\frac{1}{n}\sum_{i=1}^{n}\left(\frac{\bar{Q}_{n}^{(**)}\left(1,\,W\right)}{\psi_{1,\,n}^{(**)}}-\frac{\bar{Q}_{n}^{(**)}\left(0,\,W\right)}{\psi_{0,\,n}^{(**)}}\right)^{2}},
\end{split}
\end{equation}

where  $\bar{Q}_{n}^{(**)}$, $g_{n}^{(**)}$, $\psi_{1,\,n}^{(**)}$, and $\psi_{0,\,n}^{(**)}$ are the TMLE-updated estimates, where the targeting is done to reduce bias for the target estimand, $\Sigma^2_0$. 



\subsubsection{Initial Estimates of DGD} \label{sec:initial_est}
Let $Q_0 = (\bar{Q}_0, Q_{W, 0})$, where $Q_{W, 0}$ represents the marginal distribution of $W$. The estimand $\Sigma^2(P_0)$ is associated with $P_0$ through $Q_0$ and $g_0$. Therefore, a plug-in estimator can be constructed by estimating only the relevant parts.

The marginal distribution of covariates, $Q_{W, 0}$, is estimated using the empirical distribution of $W$, denoted as $Q_{W, n}$. We employ the SuperLearner algorithm \cite{LaanSLS2007}  to obtain the initial estimates of $\bar{Q}_0(A, W)$ and $g_0(A, W)$, denoted respectively as $\bar{Q}^{(0)}_n(A, W)$ and $g_n^{(0)}(A, W)$. SuperLearner is an ensemble machine learning framework designed to improve prediction accuracy by combining multiple base algorithms. It optimally weights these algorithms to minimize cross-validated risk, thereby producing robust and accurate predictions across diverse datasets \cite{pirracchio_improving_2015}. 

\subsubsection{Iterative TMLE}
The iterative TMLE refines initial estimates iteratively to minimize bias and ensure consistency in estimating the target parameter. Given $\Sigma^2(P)$ is a function of $g_P$ and $\bar{Q}_P$ and both require targeting, the iterative TMLE updates the initial estimates $\bar{Q}_n^{(0)}$ and $g_n^{(0)}$ during each iteration, ensuring the updated estimates $\bar{Q}_{n}^{(**)}$, $g_{n}^{(**)}$, $\psi_{1,\,n}^{(**)}$ and $\psi_{0,\,n}^{(**)}$  satisfy the EIF estimating equation: 
\begin{equation}
    P_nD^*_{\Sigma^{2}}(\bar{Q}_n^{(**)}, g_n^{(**)}, \psi_{1,\,n}^{(**)}, \psi_{0,\,n}^{(**)}) = o_P(n^{-1/2}),
\end{equation}
where $D^*_{\Sigma^2}$ is the EIF of $\Sigma^2$.  \\


\subsubsection{EIF}
To develop an efficient estimator for the variance among asymptotically linear estimators, we derive the EIF of the target parameter $\Sigma^2$ under a nonparametric model. Our approach follows the methodology of Coyle et al., who derived the variance of the EIF of the treatment-specific mean \cite{van2018targeted}. 

To simplify notation in this section and present the efficient influence function (EIF) of the target parameter more clearly, we denote $\bar{Q}_P(1, W)$ as $\bar{Q}^1$, $\bar{Q}_P(0, W)$ as $\bar{Q}^0$, $g_P(1|W)$ as $g^1$ and $g_P(0|W)$ as $g^0$. 

\begin{theorem}
    The efficient influence function of $\Sigma^2:\mathcal{M} \rightarrow \mathbb{R}$ at distribution $P \in \mathcal{M}$  is given by:
    \begin{equation} \label{eq:EIF}
    D_{\Sigma^2, P}^*(O)=D_{\Sigma^2, Q_W, P}^*(O)+D_{\Sigma^2, \bar{Q}, P}^*(O)+D_{\Sigma^2, g, P}^*(O)
\end{equation}

where
\begin{equation}
    \begin{split}
    D_{\Sigma^2, Q_W, P}^*(O) & =  \frac{1}{\left(\psi_1\right)^2} \frac{\bar{Q}^1\left(1-\bar{Q}^1\right)}{g^1}+\frac{1}{\left(\psi_0\right)^2} \frac{\bar{Q}^0\left(1-\bar{Q}^0\right)}{g^0}+\left(\frac{\bar{Q}^1}{\psi_1}-\frac{\bar{Q}^0}{\psi_0}\right)^2- \\ 
    &\quad P_W\left(\frac{1}{\left(\psi_1\right)^2} \frac{\bar{Q}^1\left(1-\bar{Q}^1\right)}{g^1}+\frac{1}{\left(\psi_0\right)^2} \frac{\bar{Q}^0\left(1-\bar{Q}^0\right)}{g^0}+\left(\frac{\bar{Q}^1}{\psi_1}-\frac{\bar{Q}^0}{\psi_1 }\right)^2\right)\\
    D_{\Sigma^2, \bar{Q}, P}^*(O) &=  \left(\frac{1}{(\psi_1)^2} \frac{A}{g^1}\left(\frac{1-2 \bar{Q}^1}{g^1}+2 \bar{Q}^1-2 \frac{\psi_1 \bar{Q}^0}{\psi_0}-\frac{2}{\psi_1}\left(P \frac{\bar{Q}^1\left(1-\bar{Q}^1\right)}{g^1}+P (\bar{Q}^1)^2\right)+\frac{2 P \bar{Q}^1 \bar{Q}^0}{\psi_0}\right)+\right. \\ 
    & \quad\left.\frac{1}{(\psi_0)^2} \frac{1-A}{g^0}\left(\frac{1-2 \bar{Q}^0}{g^0}+2 \bar{Q}^0-2 \frac{\psi_0 \bar{Q}^1}{\psi_1}-\frac{2}{\psi_0}\left(P \frac{\bar{Q}^0\left(1-\bar{Q}^0\right)}{g^0}+P (\bar{Q}^0)^2\right)+\frac{2 P \bar{Q}^1 \bar{Q}^0}{\psi_1}\right)\right)\\
    &\quad (Y-\bar{Q}(A, W)) \\
D_{\Sigma^2, g, P}^*(O) &=\left(\frac{1}{(\psi_0)^{2}}\frac{\bar{Q}^{0}\left(1-\bar{Q}^{0}\right)}{(g^{0})^{2}}\,-\,\frac{1}{(\psi_1)^{2}}\frac{\bar{Q}^{1}\left(1-\bar{Q}^{1}\right)}{(g^{1})^{2}}\,\right)\left(A-g^{1}\right)\\
\end{split}
\end{equation}

\end{theorem}
Let $T(P)$ denote the tangent space, the whole Hilbert space of mean 0, finite-variance functions, at the distribution $P$ in the model $\mathcal{M}$. Since the distribution $P$ can be factorized as $P(Y, A, W) = P(Y \mid A, W) P(A \mid W) P(W)$, the tangent space $T(P)$ is given as the orthogonal sum of the individual tangent spaces:

\begin{equation}
    T(P)= T_{Q_W}(P) \oplus T_{\bar{Q}}(P) \oplus T_{g}(P).
\end{equation}

$D_{\Sigma^2, \bar{Q}, P}^*$ is the component of the EIF in the tangent space of the conditional density of $Y$, $T_{\bar{Q}}(P)$; $D_{\Sigma^2, g, P}^*$ corresponds to the part in $T_g(P)$; $D_{\Sigma^2, Q_W, P}^*$ corresponds to the part in $T_{Q_W}(P)$. Therefore, the EIF can be orthogonally decomposed as shown in Equation \ref{eq:EIF}. The derivation of the EIF is given in Appendix \ref{appendix: aB}.


\begin{corollary}\label{corollary:EIF}
        The efficient influence function (EIF) of $\Sigma^2$ at $P$ can be represented as:
\begin{equation}
\begin{aligned}
    D_{\Sigma^2, P}^*(O) &= H_{\bar{Q}, P}(Y-\bar{Q}_P(A, W)) + H_{g, P}(A-g_P(1|W)) + D_{\Sigma^2, Q_W, P}^*, 
\end{aligned}
\end{equation}
   where the clever covariates $H_{Q, P}$ and $H_{g, P}$ are defined as:
    \begin{equation}
     \begin{aligned}
         H_{\bar{Q}, P} &= H_{1, \bar{Q}, P} + H_{0, \bar{Q}, P}\\
         H_{g, P} &=\frac{1}{\psi_0^{2}}\frac{\bar{Q}^{0}\left(1-\bar{Q}^{0}\right)}{(g^{0})^{2}}\,-\,\frac{1}{\psi_1^{2}}\frac{\bar{Q}^{1}\left(1-\bar{Q}^{1}\right)}{(g^{1})^{2}}.
     \end{aligned}
    \end{equation}

The components $H_{1, \bar{Q}, P}$ and $H_{0, \bar{Q}, P}$ are further specified as:
    \begin{equation}
        \begin{aligned}
               H_{1, \bar{Q}, P} &= \frac{1}{(\psi_1)^2} \frac{A}{g^1}\left(\frac{1-2 \bar{Q}^1}{g^1}+2 \bar{Q}^1-2 \frac{\psi_1 \bar{Q}^0}{\psi_0}-\frac{2}{\psi_1}\left(P_0 \frac{\bar{Q}^1\left(1-\bar{Q}^1\right)}{g^1}+P_0 (\bar{Q}^1)^2\right)+\frac{2 P_0 \bar{Q}^1 \bar{Q}^0}{\psi_0}\right)\\
   H_{0, \bar{Q}, P} & = \frac{1}{(\psi_0)^2} \frac{1-A}{g^0}\left(\frac{1-2 \bar{Q}^0}{g^0}+2 \bar{Q}^0-2 \frac{\psi_0 \bar{Q}^1}{\psi_1}-\frac{2}{\psi_0}\left(P_0 \frac{\bar{Q}^0\left(1-\bar{Q}^0\right)}{g^0}+P_0 (\bar{Q}^0)^2\right)+\frac{2 P_0 \bar{Q}^1 \bar{Q}^0}{\psi_1}\right).
        \end{aligned}
    \end{equation}
\end{corollary}

The EIF depends on the distribution $P$ through the treatment-specific mean $\psi_1$ and $\psi_0$, the conditional mean of outcomes $\bar{Q}_P(0, W)$ and $\bar{Q}_P(1, W)$, and the propensity function $g_P(1|W)$. With the derived EIF, we construct an asymptotically efficient estimator for the variance. The EIF characterizes both the least favorable path and the asymptotic distribution of the resulting estimator.

\begin{theorem}
    Let $O \sim P_0 \in \mathcal{M}$, and let $D^*_{\Sigma^2, P_0}$ be the efficient influence function (EIF) of $\Sigma^2$ at $P_0$. Define the second-order remainder $R_2(P_n^{**}, P_0)$ as $R\left(P_n^{**}, P_0\right)=\Sigma_{n}^{2}(P_n^{**})-\Sigma_{0}^2+P_0 D_{\Sigma^2, P_n^{**}}^*$. Suppose that the following TMLE conditions are satisfied: \begin{enumerate}
    \item Donsker class condition: $D^*_{\Sigma^2, P_n^{**}}$ belongs to a $P_0$-Donsker class with probability tending to 1. 
    \item Second-order remainder: $R_2(P_n^{**}, P_0) = o_P(n^{-\frac{1}{2}})$
    \item Consistency: $P_0\left\{D_{\Sigma^2, P_n^{**}}^*-D_{\Sigma^2, P_0}^*\right\}^2 \rightarrow_p 0$
    \end{enumerate}
    
Then the TMLE for $\Sigma^2_0$, $\Sigma^{2, TMLE}_n$, is an asymptotically efficient estimator of $\Sigma^2_0$:
\begin{equation}
    \Sigma_{n}^{2, TMLE}-\Sigma^2_{0}\,=\,\frac{1}{n} \sum_{i=1}^{n} D^*_{\Sigma^2, P_0}\left(O_{i}\right)+o_{P}\left(n^{-1 / 2}\right)
\end{equation}
\end{theorem}
The target parameter $\Sigma^2_0$ has an EIF and exact remainder that demonstrates a lack of double-robust estimation (i.e. all factors of $P_0$ have to be estimated consistently). The proof of the theorem can be found in Appendix \ref{appendix: aC}. 

\subsubsection{One-step targeted maximum likelihood estimation} \label{sec:one-step}
One-step TMLE is a refined targeting method designed to achieve efficient and unbiased parameter estimation in a single targeting step by using a universal least favorable submodel. This submodel ensures that the EIF aligns with the gradient at each parameter value, thereby avoiding iterative updates like in traditional iterative TMLE. By employing logistic ridge regression with minimal step sizes, one-step TMLE effectively mitigates overfitting, particularly in high-dimensional settings. This approach preserves the statistical properties of the initial estimator while achieving targeted bias reduction with minimal additional data fitting. Therefore, one-step TMLE provides a robust and computationally efficient alternative to iterative TMLE, especially beneficial in finite-sample settings.

Given that $Q_W$ can be estimated with its empirical distribution, we construct a universal least favorable model for $\bar{Q}$ and g. The integral representation of the universal least favorable model is: 
\begin{equation}
\begin{aligned}
{\mathrm{Logit} \bar{Q}_{\epsilon }}& ={\mathrm{Logit} \bar{Q}-\int_{0}^{\epsilon }H_{\bar{Q}, P_x} d x }\\
{\mathrm{Logit} g_{\epsilon }}& ={\mathrm{Logit} g-\int_{0}^{\epsilon }H_{g, P_x} d x ,}\\
\end{aligned}
\end{equation}

where $H_{\bar{Q}}(g, \bar{Q})$ and $H_g(g, \bar{Q})$ are the clever covariates defined in Corollary \ref{corollary:EIF}. 

The log-likelihood loss function is defined as:

\begin{equation} \label{eq:loss}
{L\left(\bar{Q},\,g\right)\,}={\,L_{1}\left(\bar{Q}\right)\left(O\right)\,+\,L_{2}\left(g\right)\left(O\right)}\\
\end{equation}
where 
\begin{equation}
    \begin{aligned}
    L_1(\bar{Q})(O) &= -\{Y \log _{}\bar{Q}(A, W)+(1-Y) \log _{}(1-\bar{Q}(A, W))\}\\
    L_2(g)(O) &= -\{A \log _{}g(1 \mid W)+(1-A) \log _{}g(0 \mid W)\}
    \end{aligned}
\end{equation}

We verify the key property of a universal least favorable model given the loss function and the definition: 

\begin{equation}
    \begin{aligned}
\frac{d}{d \varepsilon} L\left(\bar{Q}_{\varepsilon}, g_{\varepsilon}\right)&=H_{\bar{Q}, P_{\varepsilon}}\left(Y-\bar{Q}_{\varepsilon}\right)+H_{g, P_{\varepsilon}}\left(A-g_{\varepsilon}(W)\right) \\ 
&=D^*_{\Sigma^2, \bar{Q}, P_{\epsilon}}+D^*_{\Sigma^2, g, P_{\epsilon}},
\end{aligned}
\end{equation}


We describe the implementation of the one-step TMLE as follows:\\

\noindent \textbf{One-step TMLE algorithm for $\Sigma^2$ in words}
\begin{enumerate}
    \item Use SuperLearner with cross-validation to obtain initial estimates of relevant terms, including $\bar{Q}_n^{_{(0)}}(A,W), g_n^{(0)}(A|W), \psi_1^{(0)}$ and $\psi_0^{(0)}$. 
    \item  Specify the loss function as in Equation \ref{eq:loss}.
    \item Choose a small step size $d\epsilon=0.001$.
    \item     Set the iteration index $i=1$.
    \item Define the continuing criteria as follows:
\begin{itemize}
    \item $i < max\_iter$
    \item $loss_{i-1} > loss_i$
    \item $\left|P_n\left(D_{\Sigma^2, P_n}^{(i)}\right)\right| > \frac{\hat{\sigma}_{n}}{\sqrt{n} \log n}$, where $\hat{\sigma}_n$ is the standard deviation of $D^{(i)}_{\Sigma^2, P_n}$, and $n$ is the sample size. 
\end{itemize}
    \item Determine the sign of the derivative at $h=0$ of $P_nL(\bar{Q}_h, g_h)$,  which is $P_nD^*_{\Sigma^2, P_n}$. 
    \item With the initial estimates, we calculate the clever covariates $H_{\bar{Q}}$ and $H_g$ for $\bar{Q}_n^{_{(i)}}(A,W)$ and $g_n^{(i)}(A|W)$ respectively. We then update the i-th estimates with two fluctuation models:
    \begin{itemize}
        \item $\operatorname{Logit}{\bar{Q}_{\epsilon + d\epsilon}} =\operatorname{Logit} \bar{Q}_{\epsilon}-d \varepsilon H_{\bar{Q}, P_{\varepsilon}}$
        \item  $\operatorname{Logit}{g_{\epsilon + d\epsilon}} =\operatorname{Logit} g_{\epsilon}-d \varepsilon H_{g, P_{\varepsilon}} $
    \end{itemize}
\item Check the continuing criteria in Step 5 by plugging in $\bar{Q}_n^{_{(i+1)}}(A,W)$ and $g_n^{(i+1)}(A|W)$. If satisfied, return to Step 6 and continue updating. Otherwise, proceed to Step 9.
\item Plug the updated estimates $\bar{Q}_{n}^{_{(**)}}(W)$ and $g_{n}^{_{(**)}}(W)$ into Equation \ref{eq: tmle_est} to obtain the TMLE.
\end{enumerate}

\newpage

\section{Simulation}\label{sec:simulation}
In this section, we assess the robustness of the proposed targeted variance estimator through simulations in a typical causal inference setting. Specifically, we evaluate different estimators of $\Sigma^2_0$ across various sample sizes and levels of positivity. We evaluate their performance based on bias, variance and mean square error (MSE) relative to $\Sigma^2_0$, which is approximated by the Monte-Carlo variance of the TMLE estimates for log(CRR), $\Psi(P_0)$. However, our primary focus is on the coverage rates for $\Psi(P_0)$, using standard errors (SEs) derived from different variance estimators for $\Sigma^2_0$. Additional simulations exploring more complex data-generating distributions are presented in the Appendix \ref{appendix:others}.


\subsection{Simulation set-up}
We consider a point treatment setting with binary treatment and binary outcome. The data is simulated from the following data-generating distribution of $O:(W \rightarrow A \rightarrow Y) \sim P_0$:
\begin{align*}
W_{1} & \sim U(0, 1), \\
W_{2} & \sim U(0, 1), \\
W_{3} & \sim U(0, 1), \\
A &\sim Bernoulli(p), \text{where } p =  \mathrm{logit}^{-1}\left(\beta_p - (\beta_p + 2.5)W_{1} + 1.75W_{2} + (\beta_p + 3.2)W_{3}\right)\\
Y & \sim Bernoulli(q), \text{where } q = \mathrm{logit}^{-1}\left(0.1 + 0.1W_{1} + 0.1W_{2} + 0.1W_{3} + \beta_{\psi_0}A\right)
\end{align*}

Here, $\beta_p$ controls the level of positivity, ranging from -2 (no positivity violations) to 1 (severe violations), and the $\beta_{\psi_0}$ represents the effect size, with values set to 0 (null effect), 0.5 (small effect) and 2 (large effect). 

\noindent We simulate data for effect sizes $\beta_{\psi_0}$ in $\{ 0, 0.5, 2\}$, sample sizes $n$ in $\{100, 200, 500, 1000\}$, and positivity levels $\beta_p$ in $\{-2, -1.5, -1, -0.5, 0, 0.5\}$. For each combination of effect size, sample size, and positivity level, we perform 1000 iterations, generating point estimates of log(CRR) and calculating their confidence intervals for each variance estimator. To improve stability, we truncate the estimated propensity scores $g_n(A|W)$ within $[0.025, 0.975]$, and the outcome regression $\bar{Q}_n(A, W)$ within [0.001, 0.999].

Our simulations were implemented in R~\cite{R_general}, using the \texttt{tmle}  package\cite{tmle_package} and \texttt{SuperLearner} package \cite{coyle2021sl3} for estimation. Details of the implementation, including the codes that implement the proposed estimators, are available in the GitHub repository:
\url{https://github.com/wendyji220/N3C-CausalSurveillance}.\\

\subsection{Results}
\subsubsection{Results on the performance of different estimators for $\Sigma^2_0$}

\begin{figure}[H]
    \centering
    \includegraphics[width=\linewidth]{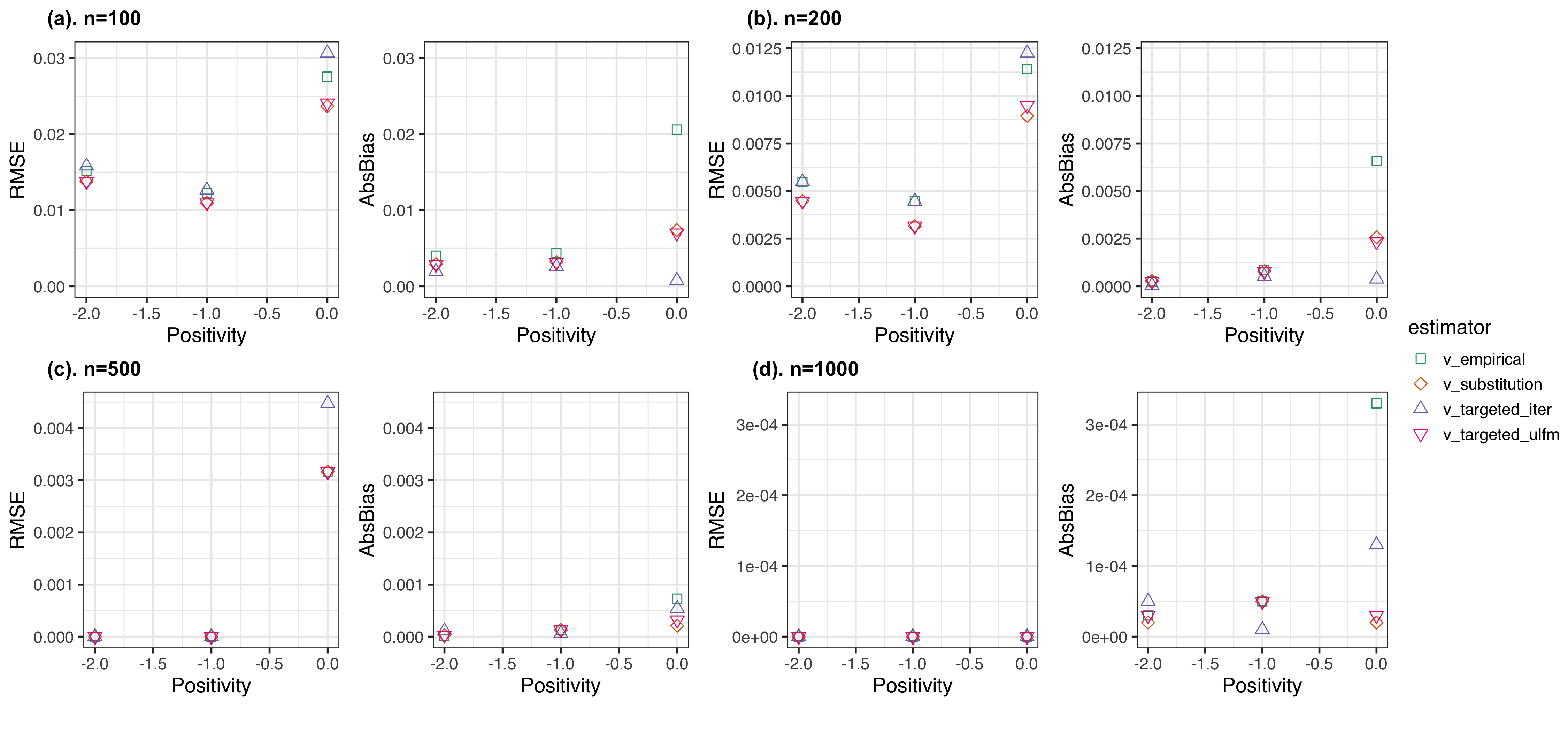}
    \caption{RMSE and absolute bias of each variance estimator compared to $\Sigma^2(P_0)$ under various sample sizes and positivity.}
    \label{p0:RMSE}
\end{figure}
Figure \ref{p0:RMSE} presents the finite-sample performance metrics for each estimator. The distribution of each variance estimator is given in Appendix \ref{appendix:dist_var}. When sample sizes are large and positivity violations are minimal, all estimators exhibit low RMSE and small bias. However, as positivity violations increase, the bias of the empirical variance estimator becomes more pronounced, leading to potential underestimation of variance. In contrast, the one-step TMLE remains stable across all settings, demonstrating robust performance. Iterative TMLE, while maintaining low bias, exhibits a higher RMSE due to increased variance, suggesting overfitting issues in finite samples.

\subsubsection{Results on the inference of log(CRR)}

\begin{figure}[H]
    \centering
    \includegraphics[width=\linewidth]{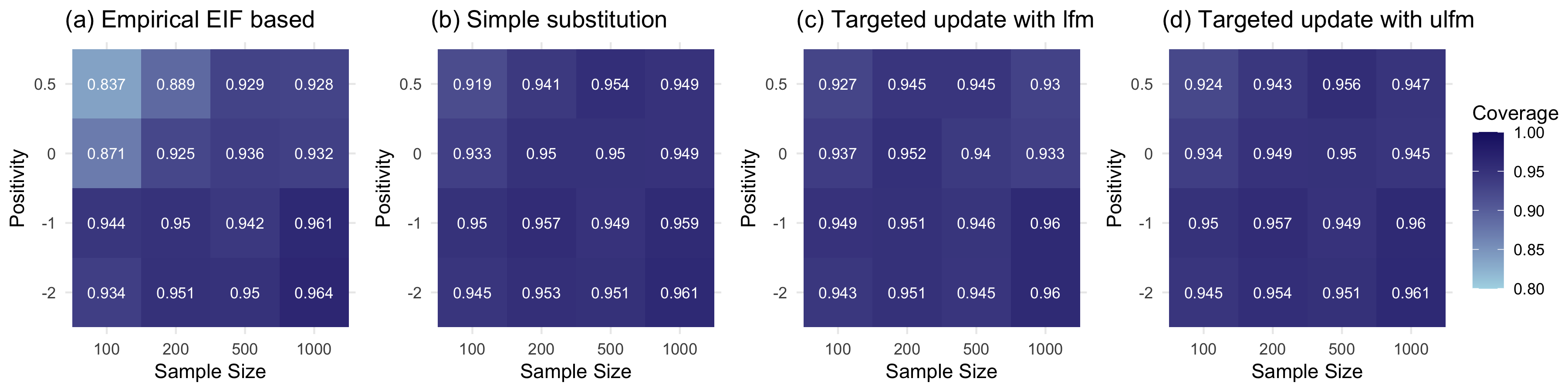}
    \caption{Coverage rates of the 95\% CIs for each variance estimator for the TMLE of log(CRR) under various sample sizes and positivity ($\beta_p$).}
    \label{p1:coverage}
\end{figure}

We present results for the case of a null effect, where $\beta_{\psi_0}=0$. Figure \ref{p1:coverage} illustrates the coverage of four variance estimators: the influence function (IF)-based estimator, the non-targeted substitution (SS) estimator, the iterative TMLE, and the one-step TMLE using the universally least favorable model. When the sample size is large and positivity is low, all estimators exhibit good coverage. However, as the sample size decreases and positivity increases, the IF-based variance estimator performs anti-conservatively, with coverage dropping below 0.85 in extreme cases. The SS estimator, iterative TMLE and one-step TMLE display similar performance, with a slight decrease to around 0.92 when the sample size is small and positivity is high. 

\begin{figure}[H]
    \centering
    \includegraphics[width=\linewidth]{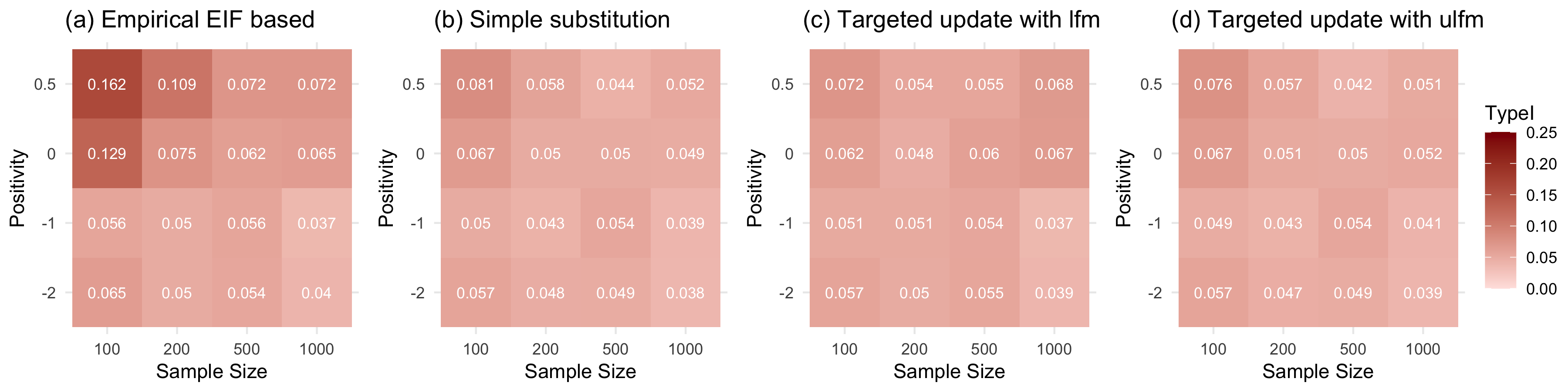}
    \caption{Type I error rates for each variance estimator for the TMLE estimator of log(CRR) under various sample sizes and positivity ($\beta_p$) when the null is true: $\Psi(P_0)=0$}
    \label{p2:Type I errors}
\end{figure}

Figure \ref{p2:Type I errors} shows the Type-I error rates for each estimator. The one-step TMLE consistently achieves a Type-I error rate near the nominal 0.05 level under the null hypothesis. In contrast, the IF-based variance estimator demonstrates elevated Type-I errors as positivity increases. 

The trends observed in these simulations are consistent across different data-generating designs (DGD). In simulations with more complex DGDs, including interaction terms, and simulations with misspecified outcome regression models, similar patterns in coverage and Type I error are observed in the Appendix \ref{appendix:others}. \\
\section{Real data-based simulation}\label{sec:data}

To evaluate the performance of the variance estimators in a real-world context, we apply them to data from the WASH Benefits Bangladesh study, which assessed the effects of interventions related to water quality, sanitation, handwashing, and nutrition on child development in rural Bangladesh  (\href{https://raw.githubusercontent.com/tlverse/tlverse-data/master/wash-benefits/washb_data.csv}{link}) \cite{Luby2018}. The WASH Benefits Bangladesh study was a large-scale, cluster-randomized controlled trial that included 4,695 pregnant women and their children, with 28 measured variables. The study aimed to evaluate the impact of four primary interventions (tr): water quality improvement using chlorine dispensers, sanitation enhancement through improved latrines and educational efforts, handwashing promotion with soap and handwashing stations, and nutritional supplementation with lipid-based nutrient supplements along with breastfeeding promotion. The primary outcome used in this analysis is the weight-for-height z-scores (WHZ), a standard measure for assessing child growth.

In this analysis, we construct a dataset similar to the simulation setting, using binary treatment and binary outcome variables. We define the treatment variable $A=\mathbb{I}(tr\neq control)$ , and the outcome variable $Y=\mathbb{I}(whz>0)$, where WHZ values greater than zero indicate healthy growth. The remaining variables serve as covariates $W$, including demographic information and other relevant factors. To ensure adequate positivity in the data, we randomly sample $n=500$ observations from the original dataset and repeat this process 10000 times. We consider a truncation level of $5/(sqrt(n)log(n))=0.036$, which is a default setting in \texttt{tmle} package, for the estimates of the propensity score $g_0(a|W)$, and calculate the proportion of the observations with truncated $g_0(a|W)$. Only samples with a truncated proportion exceeding 0.01 were selected for analysis. This process allows us to generate distributions for each of the variance estimators under certain level of positivity.

We use \texttt{tmle} package \cite{tmle_package} to compute point estimates of log(CRR) and \texttt{SuperLearner} package \cite{SL} for the initial estimation of the relevant portions of the estimand. The SuperLearner library includes a generalized linear model, a generalized linear model with penalization, a Bayesian Adaptive Regression trees model \cite{chipman2010bart} and a gradient boosting tree model \cite{chen2016xgboost}.

\begin{figure}[H]
    \centering
    \includegraphics[scale = 0.5]{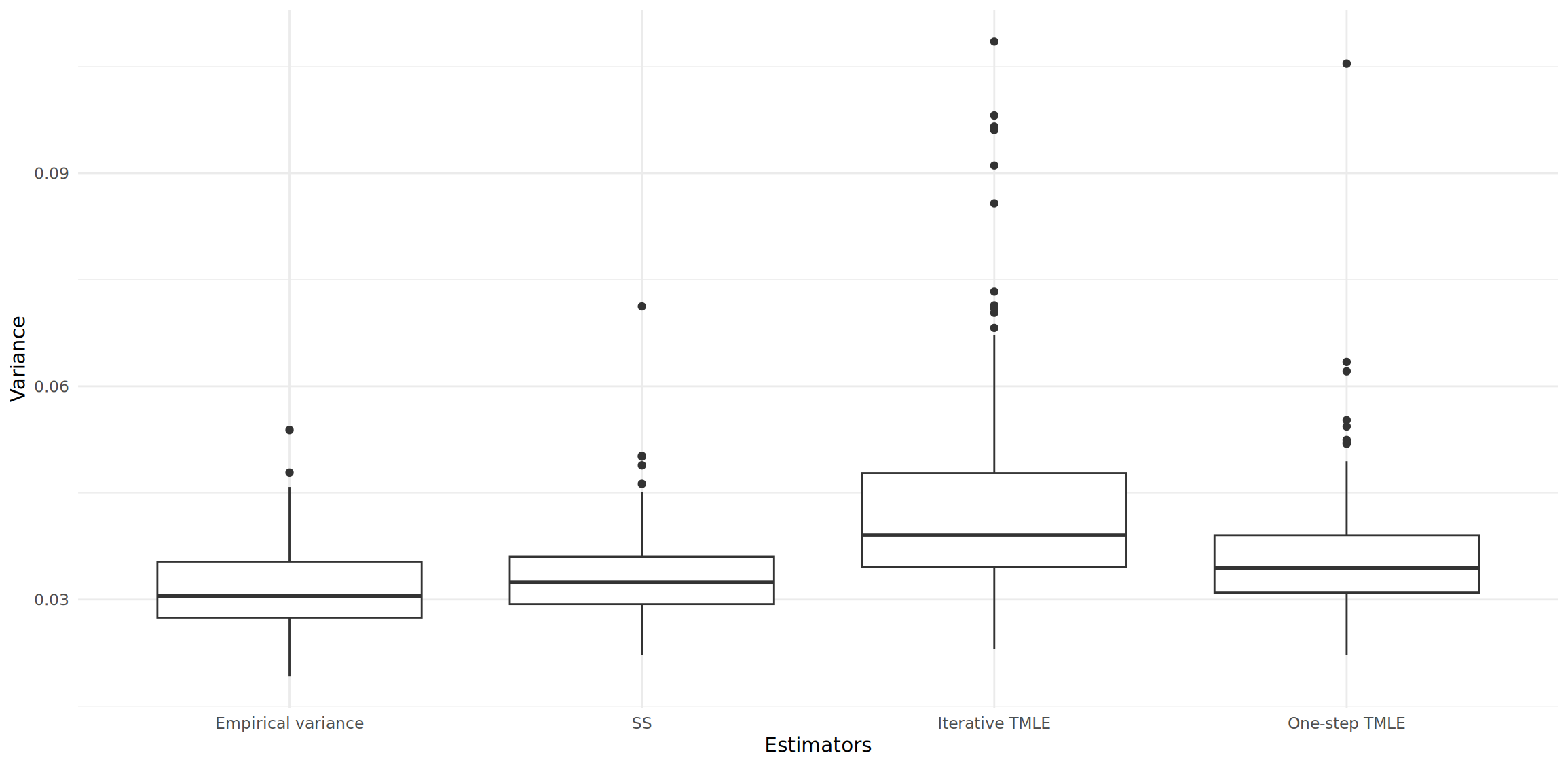}
    \caption{Distribution of estimates for each variance estimator from the WASH Benefits study.}
    \label{p4:box-plot}
\end{figure}

Figure \ref{p4:box-plot} displays the distribution of the four variance estimators based on 140 data samples from the WASH Benefits study. Consistent with the simulation results, both TMLE-based estimators provide somewhat more conservative variance estimates compared to the empirical variance under conditions of mild positivity violations. Specifically, the iterative TMLE yields systematically larger variance estimates than the one-step TMLE, indicating that iterative TMLE may be more sensitive to the complexity and nuances of real-world data. This underscores the potential advantage of the one-step TMLE in providing stable and robust variance estimation when applied to practical datasets.

\section{Discussion} \label{sec:Discussion}

In this paper, we propose a one-step Targeted Maximum Likelihood Estimator (TMLE) to estimate the variance of an asymptotically efficient estimator for the causal estimand log(CRR). Through a series of simulations, we compared the coverage rates of four variance estimators and their finite-sample performance under varying levels of positivity and sample sizes. Our results show that the coverage and Type-I errors are similar among the four estimators when the sample size is large and positivity violations are minimal. However, the one-step TMLE for variance consistently achieves nominal level coverage across all simulations, while the empirical estimator exhibits reduced coverage as positivity increases. The one-step TMLE demonstrates superior finite-sample performance compared to the iterative TMLE, as it fits the data using the universally least favorable model only once, reducing the risk of overfitting in small sample sizes. Moreover, compared to the empirical variance, the analytic form of the variance effectively captures positivity issues through terms in the denominator of Equation \ref{eq:variance}, thus avoiding underestimation and reducing bias in the presence of positivity. The application of this method to the WASH Benefits Bangladesh study demonstrated that the one-step TMLE produces more conservative estimates of variance, aligning with its robust performance in simulations. These results suggest that targeting the variance of the influence function (IF) via a universally least favorable model can lead to more reliable inference.

The structure of the variance formula, given in Equation \ref{eq:variance}, incorporates the integration of rare events, unlike the original influence function for log(CRR) presented in Equation \ref{eq:IC_logRR}, which may neglect rare events due to the indicator function of the treatment. As a result, empirical variance estimators often yield anti-conservative estimates, leading to inflated Type I error rates. 

Among the four variance estimators examined, the SS estimator, iterative TMLE, and one-step TMLE are all plug-in estimators with analytic forms. Both the iterative TMLE and one-step TMLE incorporate a crucial targeting step, where data fitting is employed to correct biases in the initial estimates by incorporating the EIF. This targeting step ensures that the resulting variance estimator is asymptotically efficient, thereby improving inference accuracy and reliability. A promising extension of this work would involve simultaneously targeting both the causal estimand and variance using a universal least favorable model, thereby obtaining a joint TMLE that is asymptotically efficient for both target parameters.

The proposed approach extends naturally to other causal estimands. For example, the iterative TMLE for the variance of IF of the treatment-specific mean has been previously introduced in Targeted Learning in Data Science \cite{van2018targeted}. The derivation of these estimators requires computing the EIF of the variance, and the step-by-step derivations provided in Appendix \ref{appendix: aB} for log(CRR) can serve as a template for other parameters. 

Future research could explore the combination of this variance estimation approach with the model-based bootstrap method proposed by Petersen et al. \cite{petersen_diagnosing_2012} to assess the severity of positivity. Such a hybrid framework could automatically select the most appropriate variance estimator, adapting to the degree of positivity issues present in the data. Parametric bootstrap offers a quantitative method for estimating bias due to positivity violations, and incorporating it within the TMLE framework could lead to an adaptive pipeline for selecting variance estimators without compromising statistical power. Additionally, the derivation of the IF for the variance estimator opens the possibility for constructing confidence intervals for the variance itself. This capability would be highly useful in practical applications, such as adaptive experiments where the upper bound of the variance estimate could be used to reduce Type I errors, thereby increasing the certainty when rejecting the null hypothesis. By refining variance estimation in this way, we can enhance the robustness of statistical inference in observational studies and randomized trials alike.

Researchers have noted that confidence intervals can become anti-conservative for complex longitudinal estimators when outcomes are rare. In particular, Nance et al. found that non-parametric bootstrap methods often provide coverage rates closer to the nominal level than empirical variance estimators of the estimated influence function \cite{nance_applying_2023}. This finding suggests a promising avenue for future research: systematically comparing various variance estimation approaches, including non-parametric bootstraps, in rare-outcome scenarios to identify those that maintain accurate coverage under these challenging conditions.

\newpage

\printbibliography

@Manual{R_general,
    title = {R: A Language and Environment for Statistical Computing},
    author = {{R Core Team}},
    organization = {R Foundation for Statistical Computing},
    address = {Vienna, Austria},
    year = {2017},
    url = {https://www.R-project.org/}
}

@manual{coyle2021sl3,
    author = {Coyle, Jeremy R and Hejazi, Nima S and Malenica, Ivana and Phillips, Rachael V and Sofrygin, Oleg},
    title = {{sl3}: Modern Pipelines for Machine Learning and {Super
    Learning}},
    year = {2021},
    howpublished = {\url{https://github.com/tlverse/sl3}},
    note = {{R} package version 1.4.2},
    url = {https://doi.org/10.5281/zenodo.1342293},
    doi = {10.5281/zenodo.1342293},
    organization = {},
    address = {}
}

@Article{tmle_package,
title = {{tmle}: An {R} Package for  Targeted Maximum Likelihood Estimation},
author = {Susan Gruber and Mark J. {van der Laan}},
journal = {Journal of Statistical Software},
year = {2012},
volume = {51},
number = {13},
pages = {1--35},
url = {https://www.jstatsoft.org/v51/i13/},
note = {doi:10.18637/jss.v051.i13},
}

@article{BenkeserHALE2016,
  author = {Benkeser, David and {van der Laan}, Mark},
  title = {The {{Highly Adaptive Lasso Estimator}}},
  year = {2016},
  journal = {IEEE International Conference on Data Science and Advanced Analytics},
  volume = {2016},
  pages = {689--696},
  doi = {10.1109/DSAA.2016.93},
  pmcid = {PMC5662030},
  pmid = {29094111}
}

@book{LaanTLCI2011,
  author = {van der Laan, Mark J. and Rose, Sherri},
  title = {Targeted {{Learning}}: {{Causal Inference}} for {{Observational}} and {{Experimental Data}}},
  year = {2011},
  shorttitle = {Targeted {{Learning}}},
  series = {Springer {{Series}} in {{Statistics}}},
  publisher = {{Springer-Verlag}},
  address = {{New York}},
  doi = {10.1007/978-1-4419-9782-1},
  isbn = {978-1-4419-9781-4},
  langid = {english}
}

@article{LaanSLS2007,
  author = {van der Laan, Mark J. and Polley, Eric C. and Hubbard, Alan E.},
  title = {Super {{Learner}}},
  year = {2007},
  journal = {Statistical Applications in Genetics and Molecular Biology},
  month = sep,
  volume = {6},
  number = {1},
  publisher = {{De Gruyter}},
  issn = {1544-6115},
  doi = {10.2202/1544-6115.1309},
  langid = {english}
}

@misc{tran_robust_2018,
	title = {Robust variance estimation and inference for causal effect estimation},
	url = {http://arxiv.org/abs/1810.03030},
	language = {en},
	urldate = {2022-10-05},
	publisher = {arXiv},
	author = {Tran, Linh and Petersen, Maya and Schwab, Joshua and van der Laan, Mark J.},
	month = oct,
	year = {2018},
	note = {arXiv:1810.03030 [math, stat]},
}

@article{petersen_diagnosing_2012,
	title = {Diagnosing and responding to violations in the positivity assumption},
	volume = {21},
	issn = {0962-2802},
	url = {https://www.ncbi.nlm.nih.gov/pmc/articles/PMC4107929/},
	doi = {10.1177/0962280210386207},
	number = {1},
	urldate = {2023-04-27},
	journal = {Statistical methods in medical research},
	author = {Petersen, Maya L and Porter, Kristin E and Gruber, Susan and Wang, Yue and van der Laan, Mark J},
	month = feb,
	year = {2012},
	pmid = {21030422},
	pmcid = {PMC4107929},
	pages = {31--54},
}

@article{van_der_laan_one-step_2016,
	title = {One-{Step} {Targeted} {Minimum} {Loss}-based {Estimation} {Based} on {Universal} {Least} {Favorable} {One}-{Dimensional} {Submodels}},
	volume = {12},
	issn = {2194-573X},
	url = {https://www.ncbi.nlm.nih.gov/pmc/articles/PMC4912007/},
	doi = {10.1515/ijb-2015-0054},
	number = {1},
	urldate = {2024-03-07},
	journal = {The international journal of biostatistics},
	author = {van der Laan, Mark and Gruber, Susan},
	month = may,
	year = {2016},
	pmid = {27227728},
	pmcid = {PMC4912007},
	pages = {351--378},
}

@book{bickel1998efficient,
  title={Efficient and Adaptive Inference in Semiparametric Models},
  author={Bickel, Peter J. and Klaassen, Chris A.J. and Ritov, Ya'acov and Wellner, Jon A.},
  year={1998},
  publisher={Springer},
  series={Springer Series in Statistics},
  address={New York},
  isbn={978-0-387-98518-3},
  doi={10.1007/978-1-4612-1842-5}
}

@book{tsiatis2006semiparametric,
  title={Semiparametric Theory and Missing Data},
  author={Tsiatis, Anastasios A.},
  year={2006},
  publisher={Springer},
  series={Springer Series in Statistics},
  address={New York},
  isbn={978-0-387-29180-2},
  doi={10.1007/0-387-36276-2}
}

@book{van2018targeted,
  title={Targeted Learning in Data Science: Causal Inference for Complex Longitudinal Studies},
  author={van der Laan, Mark J and Rose, Sherri},
  year={2018},
  publisher={Springer},
  series={Springer Series in Statistics},
  address={New York},
  isbn={978-3-319-65303-9},
  doi={10.1007/978-3-319-65304-6}
}

@article{cai2020one,
  title={One‐step targeted maximum likelihood estimation for time‐to‐event outcomes},
  author={Cai, Weixin and van der Laan, Mark J.},
  journal={Biometrics},
  volume={76},
  number={3},
  pages={722--733},
  year={2020},
  publisher={Wiley},
  doi={10.1111/biom.13172},
  url={https://doi.org/10.1111/biom.13172}
}

@article{rytgaard2024one,
  title={One-step targeted maximum likelihood estimation for targeting cause-specific absolute risks and survival curves},
  author={Rytgaard, H. C. W. and van der Laan, M. J.},
  journal={Biometrika},
  volume={111},
  number={1},
  pages={129--145},
  year={2024},
  publisher={Oxford University Press},
  doi={10.1093/biomet/asad033},
  url={https://doi.org/10.1093/biomet/asad033}
}

@article{Luby2018,
  title={Effects of Water Quality, Sanitation, Handwashing, and Nutritional Interventions on Diarrhoea and Child Growth in Rural Bangladesh: A Cluster Randomised Controlled Trial},
  author={Luby, Stephen P and Rahman, Mahbubur and Arnold, Benjamin F and Unicomb, Leanne and Ashraf, Sania and Winch, Peter J and Stewart, Christine P and others},
  journal={The Lancet Global Health},
  volume={6},
  number={3},
  pages={e302--e315},
  year={2018},
  publisher={Elsevier}
}

@Manual{SL,
  title = {SuperLearner: A prediction algorithm for combining models},
  author = {Eric C. Polley and Alan E. Hubbard},
  year = {2019},
  note = {R package version 2.0-26},
  url = {https://CRAN.R-project.org/package=SuperLearner},
}

@inproceedings{chen2016xgboost,
  title={XGBoost: A Scalable Tree Boosting System},
  author={Chen, Tianqi and Guestrin, Carlos},
  booktitle={Proceedings of the 22nd ACM SIGKDD International Conference on Knowledge Discovery and Data Mining},
  pages={785--794},
  year={2016},
  organization={ACM}
}

@article{chipman2010bart,
  title={BART: Bayesian Additive Regression Trees},
  author={Chipman, Hugh A and George, Edward I and McCulloch, Robert E},
  journal={The Annals of Applied Statistics},
  volume={4},
  number={1},
  pages={266--298},
  year={2010},
  publisher={Institute of Mathematical Statistics}
}

@article{vanderLaanRubin2006,
  title={Targeted maximum likelihood learning},
  author={van der Laan, Mark J. and Rubin, Daniel},
  journal={The International Journal of Biostatistics},
  volume={2},
  number={1},
  year={2006},
  pages={1--40},
  doi={10.2202/1557-4679.1043}
}

@article{vanderLaanGruber2011,
  title={Targeted Minimum Loss Based Estimation of an Intervention Specific Mean Outcome},
  author={van der Laan, Mark J. and Gruber, Susan},
  journal={The Berkeley Electronic Press},
  year={2011},
  number={290},
  url={https://biostats.bepress.com/ucbbiostat/paper290/}
}

@book{pearl2000causality,
  title        = {Causality: Models, Reasoning and Inference},
  author       = {Judea Pearl},
  year         = {2000},
  publisher    = {Cambridge University Press},
  address      = {Cambridge, UK},
  edition      = {1st},
  volume       = {19},
  number       = {2},
  pages        = {3}
}

@article{vonmises1947,
  author    = {Richard von Mises},
  title     = {On the asymptotic distribution of differentiable statistical functions},
  journal   = {The Annals of Mathematical Statistics},
  volume    = {18},
  number    = {3},
  pages     = {309--348},
  year      = {1947},
  publisher = {Institute of Mathematical Statistics},
  doi       = {10.1214/aoms/1177730380},
  url       = {https://projecteuclid.org/euclid.aoms/1177730380}
}

@article{li_evaluating_2022,
	title = {Evaluating the robustness of targeted maximum likelihood estimators via realistic simulations in nutrition intervention trials},
	volume = {41},
	issn = {0277-6715},
	url = {https://www.ncbi.nlm.nih.gov/pmc/articles/PMC10362909/},
	doi = {10.1002/sim.9348},
	number = {12},
	urldate = {2023-08-17},
	journal = {Statistics in medicine},
	author = {Li, Haodong and Rosete, Sonali and Coyle, Jeremy and Phillips, Rachael V. and Hejazi, Nima S. and Malenica, Ivana and Arnold, Benjamin F. and Benjamin-Chung, Jade and Mertens, Andrew and Colford, John M. and van der Laan, Mark J. and Hubbard, Alan E.},
	month = may,
	year = {2022},
	pmid = {35172378},
	pmcid = {PMC10362909},
	pages = {2132--2165}
}

@article{pirracchio_improving_2015,
	title = {Improving {Propensity} {Score} {Estimators}' {Robustness} to {Model} {Misspecification} {Using} {Super} {Learner}},
	volume = {181},
	issn = {0002-9262},
	url = {https://www.ncbi.nlm.nih.gov/pmc/articles/PMC4351345/},
	doi = {10.1093/aje/kwu253},
	number = {2},
	urldate = {2025-03-06},
	journal = {American Journal of Epidemiology},
	author = {Pirracchio, Romain and Petersen, Maya L. and van der Laan, Mark},
	month = jan,
	year = {2015},
	pmid = {25515168},
	pmcid = {PMC4351345},
	pages = {108--119}
}

@misc{nance_applying_2023,
	title = {Applying the causal roadmap to longitudinal national {Danish} registry data: a case study of second-line diabetes medication and dementia},
	shorttitle = {Applying the causal roadmap to longitudinal national {Danish} registry data},
	url = {http://arxiv.org/abs/2310.03235},
	doi = {10.48550/arXiv.2310.03235},
	urldate = {2025-03-06},
	publisher = {arXiv},
	author = {Nance, Nerissa and Mertens, Andrew and Gerds, Thomas and Wang, Zeyi and Torp-Pedersen, Christian and Laan, Mark van der and Kvist, Kajsa and Lange, Theis and Zareini, Bochra and Petersen, Maya},
	month = oct,
	year = {2023},
	note = {arXiv:2310.03235 [stat]}
}
\newpage
\appendix

\section{Derivation of target parameter - variance of IC of log(CRR)} \label{appendix: aA}

We derive the analytic form of the variance of the Efficient Influence Function (EIF) for the log of the Causal Risk Ratio (log(CRR)). The EIF for log(CRR) is given in Equation \ref{eq:IC_logRR}. To obtain the variance of this EIF, we simply calculate the variance of the expression in Equation \ref{eq:IC_logRR}.
\begin{equation}
\begin{split}
{\Sigma^{2}\left(P_{0}\right)}& ={\mathbb{E}_{0}[D^{*}_{\Psi, P_0}(O)]^{2}}\\
&={\mathbb{E}_{0}\bigg[\bigg(\frac{\mathbb{I}(A=1)}{\mathbb{E}Y^{1}g_{0}(1 \mid W)}-\frac{\mathbb{I}(A=0)}{\mathbb{E}Y^{0}g_{0}(0 \mid W)}\bigg)\left(Y-\bar{Q}_{0}(A, W)\right) + \frac{\bar{Q}_{0}(1, W)}{\mathbb{E}Y^{1}} - \frac{\bar{Q}_{0}(0, W)}{\mathbb{E}Y^{0}}\bigg]^{2}}\\
{ }&={\mathbb{E}_{0}\bigg[\bigg(\frac{\mathbb{I}(A=1)}{\mathbb{E}Y^{1}g_{0}(1 \mid W)}-\frac{\mathbb{I}(A=0)}{\mathbb{E}Y^{0}g_{0}(0 \mid W)}\bigg)\left(Y-\bar{Q}_{0}(A, W)\right)\bigg]^{2} + \,\mathbb{E}_{0}\bigg[\frac{\bar{Q}_{0}(1, W)}{\mathbb{E}Y^{1}} - \frac{\bar{Q}_{0}(0, W)}{\mathbb{E}Y^{0}}\bigg]^{2}}\\
{ }&={\mathbb{E}_{0}\bigg[\bigg(\frac{\mathbb{I}(A=1)}{\mathbb{E}Y^{1}g_{0}(1 \mid W)}-\frac{\mathbb{I}(A=0)}{\mathbb{E}Y^{0}g_{0}(0 \mid W)}\bigg)^{2}\left(Y-\bar{Q}_{0}(A, W)\right)^{2}\bigg] + \,...}\\
{}&={\mathbb{E}_{0}\bigg[\frac{(Y^{1}-\bar{Q}_{0}(1, W))^{2}}{(\mathbb{E}Y^{1}g_{0}(1 \mid W))^{2}}+\frac{(Y^{0}-\bar{Q}_{0}(0, W))^{2}}{(\mathbb{E}Y^{0}g_{0}(0 \mid W))^{2}}\bigg] + \,...}\\
{\,}&=\frac{1}{(\mathbb{E}Y^{1})^{2}}\mathbb{E}_{P_{0}^{1}}\bigg[\frac{(Y^{1}-\bar{Q}_{0}(1, W))^{2}}{g_{0}(1 \mid W)}\bigg] +\frac{1}{(\mathbb{E}Y^{0})^{2}} \mathbb{E}_{P_{0}^{0}}\bigg[\frac{(Y^{0}-\bar{Q}_{0}(0, W))^{2}}{g_{0}(0 \mid W)}\bigg] \\
&\quad + \mathbb{E}_{0}\bigg[\frac{\bar{Q}_{0}(1, W)}{\mathbb{E}Y^{1}} - \frac{\bar{Q}_{0}(0, W)}{\mathbb{E}Y^{0}}\bigg]^{2}\\
&=\frac{1}{(\mathbb{E}Y^{1})^{2}}\mathbb{E}_{P_{0}^{1}}\bigg[\frac{(\bar{Q}_{0}(1, W)\left(1-\bar{Q}_{0}(1, W)\right)}{g_{0}(1 \mid W)}\bigg] +\frac{1}{(\mathbb{E}Y^{0})^{2}} \mathbb{E}_{P_{0}^{0}}\bigg[\frac{(\bar{Q}_{0}(0, W)\left(1-\bar{Q}_{0}(0, W)\right)}{g_{0}(0 \mid W)}\bigg]\\
&\quad +\mathbb{E}_{0}\bigg[\frac{\bar{Q}_{0}(1, W)}{\mathbb{E}Y^{1}} - \frac{\bar{Q}_{0}(0, W)}{\mathbb{E}Y^{0}}\bigg]^{2}
\end{split}
\end{equation}

It gives us the representation of the target parameter in Equation \ref{eq:variance}, which is:
\begin{equation}
\begin{split}
    \Sigma^2(P)  & = P_{W, 0}\left(\frac{1}{\left(\mathbb{E} Y^1\right)^2} \frac{\left(\bar{Q}_0(1, W)\left(1-\bar{Q}_0(1, W)\right)\right)}{g_0(1 \mid W)}+\frac{1}{\left(\mathbb{E} Y^0\right)^2} \frac{\left(\bar{Q}_0(0, W)\left(1-\bar{Q}_0(0, W)\right)\right)}{g_0(0 \mid W)}\right.\\
    &\quad+\left.\left(\frac{\bar{Q}_0(1, W)}{\mathbb{E} Y^1}-\frac{\bar{Q}_0(0, W)}{\mathbb{E} Y^0}\right)^2\right).
\end{split}
\end{equation}

\section{Derivation of the efficient Influence function (EIF) of the target parameter, $\Sigma^2_0$} \label{appendix: aB}

To construct a TMLE or other asymptotically linear estimators, we first derive the efficient influence function (EIF) for the target parameter. The EIF, denoted as $D_{\Sigma^2}^*(P)$, can be decomposed as projections onto three tangent spaces.

For simplicity of notation, we introduce abbreviated terms for the components in Equation\ref{eq:variance}. We denote $\mathbb{E} Y^1$ and $\mathbb{E} Y^0$ as $\mu_1$ and $\mu_0$, respectively. We then define the following quantities:
\begin{equation}
\begin{aligned}
        \theta_1 &= P_0\frac{\bar{Q}_0(1, W)\left(1-\bar{Q}_0(1, W)\right)}{g_0(1 \mid W)}\\
    \theta_2 &= P_0\frac{\bar{Q}_0(0, W)\left(1-\bar{Q}_0(0, W)\right)}{g_0(0 \mid W)}\\
    \theta_3 &= P_0 \bar{Q}_0(1, W)^2\\
    \theta_4 &= P_0 \bar{Q}_0(0, W)^2 \\
    \theta_5 &= P_0 \bar{Q}_0(1, W)Q_1(0, W) \\
\end{aligned}
\end{equation}

Using these simplified notations, we can express Equation \ref{eq:variance} as follows:

\begin{equation} \label{eq:simple_var}
    \Sigma^2(P) = \frac{1}{\mu_1^2} \theta_1+\frac{1}{\mu_0^2} \theta_2+\frac{1}{\mu_1^2} \theta_3+\frac{1}{\mu_0^2} \theta_4-\frac{2}{\mu_1 \mu_0} \theta_5
\end{equation}

The EIF of $\Sigma^2(P)$ can then be derived using the delta method and the EIFs of the individual components in Equation \ref{eq:simple_var}:
\begin{equation}
\begin{split}
        D_{\Sigma}^*(P) =& \left(-\frac{2}{\mu_1^3}\left(\theta_1+\theta_3\right)+\frac{2 \theta_5}{\mu_1^2 \mu_0}\right) D_{\mu_1}^*+\left(-\frac{2}{\mu_0^3}\left(\theta_2+\theta_4\right)+\frac{2 \theta_5}{\mu_0^2 \mu_1}\right) D_{\mu_0}^*\\
       & + \frac{1}{\mu_1^2} D_{\theta_1}^* + \frac{1}{\mu_0^2} D_{\theta_2}^* + \frac{1}{\mu_1^2} D_{\theta_3}^* + \frac{1}{\mu_0^2} D_{\theta_4}^* -\frac{2}{\mu_1 \mu_0} D_{\theta_5}^*
\end{split}
\end{equation}
We obtained the EIF of each components using the method presented in A.3 in the targeted learning book \cite{vanderLaanGruber2011} by considering a non-parametric estimator and assuming the data structure is discrete. Combining these component‑specific EIFs yields the efficient influence function for our target parameter, as presented in Equation \ref{eq:EIF}.

\section{Theorem 2}  \label{appendix: aC}
We aim to prove the asymptotic efficiency of the TMLE for $\Sigma^2$, specifically: 
\begin{equation} \label{efficiency}
    \Sigma^{2. TMLE}_n - \Sigma^2_0 = P_n D^*_{P_0} + o_P(n^{-1/2})
\end{equation}

We consider an asymptotically linear estimator at $P_n$ with an influence function $D^*(P_n)$. Using the von Mises expansion of $\Sigma^2$ centered at $P_0$ \cite{vonmises1947} and defining $R\left(P_n, P_0\right)=\Sigma^{2}(P_n)-\Sigma^2(P_0)+P_0 D_{P_n}^* $, we obtain the exact second order expansion of $\Sigma^2$ as:
\begin{equation}
    \begin{aligned}
        \Sigma^2_n - \Sigma^2_0 &= -P_0D^*_{P_n} + R(P_n, P_0). \\
    \end{aligned}
\end{equation}

Applying algebraic manipulations and rearranging terms, we express the expansion as:
\begin{equation}
\begin{aligned}
    \Sigma^2_n - \Sigma^2_0 &= -P_0D^*_{P_n} + R(P_n, P_0)\\
& =\left(P_n-P_0\right)\left(D_{P_n}^*-D_{P_0}^*\right)+P_n D_{P_0}^*-P_n D_{P_n}^*+R\left(P_n, P_0\right) \\
&= P_n D_{P_0}^*+\left(P_n-P_0\right)\left(D_{P_n}^*-D_{P_0}^*\right)-P_n D_{P_n}^*+R\left(P_n, P_0\right)
\end{aligned}
\end{equation}
For the estimator to achieve efficiency, the last three terms must be $o_P(n^{-1/2})$. Each term is controlled as follows:
\begin{itemize}
    \item $(P_n-P_0)(D_{P_n}^*-D_{P_0}^*)$: This term is $o_P(n^{-1/2})$ under the Donsker condition (A1) and L1-consistency condition (A3).
    \item $P_nD^*_{P_n}$: This term is controlled by the TMLE targeting step, ensuring it is $o_P(n^{-1/2})$.
    \item $R(P_n, P_0)$: This remainder term is $o_P(n^{-1/2})$ under Assumption 2, which we analyze in detail below.  
\end{itemize}

\subsection{Second-order remainder}
The target parameter in Equation \ref{eq:variance} can be expressed as a combination of five distinct statistical components:

\begin{equation}
    \begin{aligned}
        \Sigma^2(P) = \frac{1}{(\mu_1)^2}\Sigma^{2, \psi_1}(P) + \frac{1}{(\mu_0)^2}\Sigma^{2, \psi_0}(P) - \frac{2}{\mu_1\mu_0} \mathbb{E}D^{*}_{\psi_1, P} D^{*}_{\psi_0, P},
    \end{aligned}
\end{equation}
where $\Sigma^{2, \psi_a}(P)$ is the variance of EIF for $\psi_a$ at distribution $P$. 

Letting $\sigma(P)$ represent $\mathbb{E}D^{*}_{\psi_1, P} D^{*}_{\psi_0, P}$, we can apply the delta method to derive the EIF of $\Sigma^2(P)$ as follows:
\begin{equation}
    \begin{aligned}
        D^*_{\Sigma^2, P} = \frac{1}{(\mu_1)^2}D^*_{\Sigma^{2, \psi_1},P} + \frac{1}{(\mu_0)^2}D^*_{\Sigma^{2, \psi_0}, P} - \frac{2}{\mu_1\mu_0} D^*_{\sigma, P} +aD^*_{\mu_1, P} + bD^*_{\mu_0, P},
    \end{aligned}
\end{equation}
where $a$ and $b$ are are coefficients obtained from the delta method expansion. 

Therefore, we can write the second-order remainder of the target parameter as:
\begin{equation}
    \begin{split}
        R(P_n, P_0) &= \Sigma^{2}(P_n)-\Sigma^{2}(P_0) + P_0D^*_{\Sigma^2, P_n}\\
        &=\Sigma^{2}(P_n)-\Sigma^{2}(P_0) + \frac{1}{(\mu_1)^2}P_0D_{\Sigma^{2, \psi_1},P_n} + \frac{1}{(\mu_0)^2}P_0D_{\Sigma^{2, \psi_0}, P_n} - \frac{2}{\mu_1\mu_0} P_0D^*_{\sigma, P_n} \\
        & \quad+ aP_0D^*_{\mu_1, P_n} + bP_0D^*_{\mu_0, P_n}
    \end{split}
\end{equation}

Let $\bm{\theta} (P)= (\Sigma^{2, \psi_1}(P_n), \Sigma^{2, \psi_0}(P_n), \sigma(P_n), \mu_1, \mu_0)$. By the Taylor expansion for multivariate functions, we rewrite $\Sigma^2(P_n)-\Sigma^2(P_0)$ as:

\begin{equation}
    \begin{split}
        \Sigma^2(P_n)-\Sigma^2(P_0)& = f(\bm{\theta}(P_n)) - f(\bm{\theta}(P_0))\\
        &=\sum _j  \frac{df(\bm{\theta}(P_0))}{d\theta_j}(\theta_j(P_n) - \theta_{j}(P_0)) + R(\bm{\theta}(P_n), \bm{\theta}(P_0))
    \end{split}
\end{equation}
Therefore,
\begin{equation}\label{eq:remainder_linear_comb}
    \begin{split}
        R(P_n, P_0) &= \sum _j  \frac{df(\bm{\theta}(P_0))}{d\theta_j}(\theta_j(P_n) - \theta_{j}(P_0)+P_0 D^*_{\theta_j,P_n}) + R(\bm{\theta}(P_n), \bm{\theta}(P_0))\\
        & =  \sum _j  \frac{df(\bm{\theta}(P_0))}{d\theta_j}R_{\theta_j}(P_n, P_0) +R(\bm{\theta}(P_n), \bm{\theta}(P_0))
    \end{split}
\end{equation}

The second-order remainder is a linear combination of the exact remainders of $\mu_1, \mu_0, \Sigma^{2, \psi_1}(P), \Sigma^{2, \psi_0}(P)$ and $\sigma(P)$, which we can derive in order to get that of $\Sigma^2(P)$. The remainders for $\mu_1$ and $\mu_0$ are already known to be doubly robust \cite{van2018targeted}. Consequently, the main task is to derive the exact remainders for $\Sigma^{2, \psi_1}, \Sigma^{2, \psi_0}$ and $\sigma$; these derivations are provided in the sections that follow.

\subsubsection{Second-order remainder of $\Sigma^{2, \psi_a}(P)$}

 The variance $\Sigma^{2, \psi_a}$ at distribution $P$ is given by:
\begin{equation}
\Sigma^{2, \psi_a}\left(P\right)  =E_W\left(\frac{\bar{Q}_P(a, W)\left(1-\bar{Q}_P(a, W)\right)}{g_P(a\mid W)}+\left(\bar{Q}_P(a, W)-\psi_a\right)^2\right)
\end{equation}

The efficient influence function $D_{\Sigma, \psi_a}^*(P)$ of $\Sigma^{2, \psi_a}$ at $P$ is \cite{van2018targeted}:

$$
D_{\Sigma, \psi_a}^*(P)(W, A, Y)=D_{\Sigma^2, \psi_a, Q_W}(P)(W)+D_{\Sigma^2, \psi_a, \bar{Q}}(P)(O)+D_{\Sigma^2,\psi_a,  g}(P)(O)
$$
where
$$
\begin{aligned}
D_{\Sigma^2, \psi_a, Q_W}(P)(W)= & \frac{\bar{Q}^a(1-\bar{Q}^a)}{g^a}(W)-Q_W \frac{\bar{Q}^a(1-\bar{Q}^a)}{g^a} +(\bar{Q}^a-\psi_a)^2-P_W(\bar{Q}^a-\psi_a)^2 \\
D_{\Sigma^2, \psi_a, \bar{Q}}(P)(O)= & \frac{I(A=a)}{g^a}\left(\frac{1-2 \bar{Q}^a}{g^a}+2(\bar{Q}^a-\psi_a)\right)(Y-\bar{Q}^a) \\
D_{\Sigma^2, \psi_a, g}(P)(O)= & -\frac{\bar{Q}^a(1-\bar{Q}^a)}{(g^a)^2}(A-g^a).
\end{aligned}
$$

Let $\bar{Q}_0 \equiv \bar{Q}_{P_0}(a, W)$ and $\bar{Q}_n \equiv \bar{Q}_{P_n}(a, W)$. Similarly, $g_0 \equiv g_{P_0}(a\mid W)$ and $g_n \equiv g_{P_n}(a\mid W)$. We write the second-order remainder as:
\begin{equation}
    \begin{aligned}
        R(P_n, P_0)  = &\Sigma^{2, \psi_1}(P_n)-\Sigma^{2, \psi_1}(P_0) + P_0D^*_{\Sigma^2, P_n} \\   
    = &{\Sigma^{2, \psi_1}(P_n)-\Sigma^{2, \psi_1}\left(P_{0}\right)+P_{0} D_{\Sigma^{2}, Q_{W}}(P_n)+P_{0} D_{\Sigma^{2}, \bar{Q}}(P_n)+P_{0} D_{\Sigma^{2}, g_n}(P_n)
}\\
=&{-P_{0}\left(\frac{\bar{Q}_0\left(1-\bar{Q}_0\right)}{g_{0}}+\left(\bar{Q}_0-P_{0}\bar{Q}_0\right)^{2}\right)}+P_{0} \frac{\bar{Q}_n(1-\bar{Q}_n)}{g}+P_{0}\left(\bar{Q}_n-P\bar{Q}_n\right)^{2}+\\
&P_{0} \frac{A}{g_n}\left(\frac{1-2 \bar{Q}_n}{g_n}+2\left(\bar{Q}_n-P_n\bar{Q}_n\right)\right)(Y-\bar{Q}_n)-P_{0} \frac{\bar{Q}_n(1-\bar{Q}_n)}{g^{2}}(A-g_n\,)\\
=&{-P_{0}\left(\frac{\bar{Q}_0\left(1-\bar{Q}_0\right)}{g_{0}}+\left(\bar{Q}_0-P_{0}\bar{Q}_0\right)^{2}\right)}+P_{0} \frac{\bar{Q}_n(1-\bar{Q}_n)}{g}+P_{0}\left(\bar{Q}_n-P_n\bar{Q}_n\right)^{2}+\\
&P_{0} \left\{P_0\left[\frac{A}{g_n}\left(\frac{1-2 \bar{Q}_n}{g_n}+2\left(\bar{Q}_n-P_n\bar{Q}_n\right)\right)(Y-\bar{Q}_n)\mid W\right]\right\}
-\\
&P_{0} \left\{P_0\left[\frac{\bar{Q}_n(1-\bar{Q}_n)}{(g_n)^{2}}(A-g_n\,)\mid W\right]\right\}\\
=&{-P_{0}\left(\frac{\bar{Q}_0\left(1-\bar{Q}_0\right)}{g_{0}}+\left(\bar{Q}_0-P_{0}\bar{Q}_0\right)^{2}\right)}+P_{0} \frac{\bar{Q}_n(1-\bar{Q}_n)}{g_n}+P_{0}\left(\bar{Q}_n-P\bar{Q}_n\right)^{2}+\\
&P_{0} \frac{g_0}{g_n}\left(\frac{1-2 \bar{Q}_n}{g_n}+2\left(\bar{Q}_n-P_n\bar{Q}_n\right)\right)(\bar{Q}_0-\bar{Q}_n)-P_{0} \frac{\bar{Q}_n(1-\bar{Q}_n)}{(g_n)^{2}}(g_0-g_n\,)\\
=&-P_{0}\frac{\bar{Q}_0\left(1-\bar{Q}_0\right)}{g_{0}}-P_0\left(\bar{Q}_0-P_{0}\bar{Q}_0\right)^{2}
+P_{0} \frac{\bar{Q}_n(1-\bar{Q}_n)}{g_n}+P_{0}\left(\bar{Q}_n-P_n\bar{Q}_n\right)^{2}+\\
&P_{0} \left(\frac{g_0}{g_n}-1\right)\left(\frac{1-2 \bar{Q}_n}{g_n}+2\left(\bar{Q}_n-P_n\bar{Q}_n\right)\right)(\bar{Q}_0-\bar{Q}_n)
-P_{0} \frac{\bar{Q}_n(1-\bar{Q}_n)}{(g_n)^{2}}(g_0-g_n\,) +\\
&P_0 \left(\frac{1-2 \bar{Q}_n}{g_n}+2\left(\bar{Q}_n-P_n\bar{Q}_n\right)\right)(\bar{Q}_0-\bar{Q}_n)
    \end{aligned}
\end{equation}

We need to do some rearrangement of the terms. We denote each term as:

\begin{equation}
\begin{aligned}
  \theta_1 &=-P_{0}\frac{\bar{Q}_0\left(1-\bar{Q}_0\right)}{g_{0}}\\
    \theta_2 &= -P_0\left(\bar{Q}_0-P_{0}\bar{Q}_0\right)^{2}\\
    \theta_3 &= P_{0} \frac{\bar{Q}_n(1-\bar{Q}_n)}{g_n}\\
    \theta_4 &= P_{0}\left(\bar{Q}_n-P_n\bar{Q}_n\right)^{2} \\
    \theta_5 &= P_{0} (\frac{g_0}{g_n}-1)\left(\frac{1-2 \bar{Q}_n}{g_n}+2\left(\bar{Q}_n-P_n\bar{Q}_n\right)\right)(\bar{Q}_0-\bar{Q}_n)\\
    \theta_6 &=-P_{0} \frac{\bar{Q}_n(1-\bar{Q}_n)}{(g_n)^{2}}(g_0-g_n\,)\\
    \theta_7 &= P_0 \left(\frac{1-2 \bar{Q}_n}{g_n}\right)(\bar{Q}_0-\bar{Q}_n)\\
    \theta_8 &= P_0 2\left(\bar{Q}_n-P_n\bar{Q}_n\right)(\bar{Q}_0-\bar{Q}_n)
\end{aligned}
\end{equation}
Note that $\theta_5$ already contains the product of $g_0-g$ and $\bar{Q}_0-Q$, which is a standard double robust term. Therefore, we focus on the remaining 7 terms through algebraic manipulations and grouping of terms.
\begin{equation}
    \begin{aligned}
        \theta_2 + \theta_4 + \theta_8 &= -P_0\left(\bar{Q}_0-P_{0}\bar{Q}_0\right)^{2} + P_{0}\left(\bar{Q}_n-P\bar{Q}_n\right)^{2} + P_0 2\left(\bar{Q}_n-P\bar{Q}_n\right)(\bar{Q}_0-\bar{Q}_n) \\
        & = P_0\left \{(\bar{Q}_n-P \bar{Q}_n)\left(2 \bar{Q}_0-\bar{Q}_n-P \bar{Q}_n\right)-\left(\bar{Q}_0-P_0 \bar{Q}_0\right)^2\right\}\\
        &= -P_0\left(\bar{Q}_n-\bar{Q}_0\right)^2+\left(P_n \bar{Q}_n-P_0 \bar{Q}_0\right)^2
    \end{aligned}
\end{equation}

\begin{equation}
    \begin{split}
        \theta_1 + \theta_3 + \theta_6 + \theta_7 &= -P_0\left(\frac{\bar{Q}_0\left(1-\bar{Q}_0\right)}{g_0}\right)+P_0\left(\frac{\bar{Q}_n(1-\bar{Q}_n)}{g}\right)-P_0\left[\bar{Q}_n(1-\bar{Q}_n) \frac{g_0-g_n}{(g_n)^2}\right]\\
        &\quad+P_0\left(\frac{1-2 \bar{Q}_n}{g_n}\right)\left(\bar{Q}_0-Q\right) \\
        &= P_0\left\{-\frac{\bar{Q}_0}{g_0}+\frac{(\bar{Q}_0)^2}{g_0}-\bar{Q}_n(1-\bar{Q}_n) \frac{g_0-g}{(g_n)^2}+\frac{\bar{Q}_0}{g_n}-\frac{2 \bar{Q}_n \bar{Q}_0}{g_n}+\frac{(\bar{Q}_n)^2}{g_n}\right\}\\
        &= P_0\left\{-\frac{\bar{Q}_0}{g_0}+\frac{(\bar{Q}_0)^2}{g_0}-\bar{Q}_n(1-\bar{Q}_n) \frac{g_0-g_n}{(g_n)^2}+\frac{\bar{Q}_0}{g_n}-\frac{2 \bar{Q}_n \bar{Q}_0}{g_n}+\frac{(\bar{Q}_n)^2}{g_n}+\frac{\bar{Q}_0^2}{g_n}-\frac{(\bar{Q}_0)^2}{g_n}\right\}\\
        & = P_0\left\{\left(\frac{\bar{Q}_0}{g_n}-\frac{\bar{Q}_0}{g_0}\right)+\left(\frac{\bar{Q}_0^2}{g_0}-\frac{(\bar{Q}_0)^2}{g_n}\right)-\bar{Q}_n(1-\bar{Q}_n) \frac{g_0-g}{g^2}+\frac{\left(\bar{Q}_n-\bar{Q}_0\right)^2}{g_n}\right\}\\
        &= P_0\left\{\bar{Q}_0 \frac{g_0-g_n}{g_n g_0}-\bar{Q}_0^2 \frac{g_0-g_n}{g_n g_0}-\bar{Q}_n(1-\bar{Q}_n) \frac{g_0-g_n}{(g_n)^2}+\frac{\left(\bar{Q}_n-\bar{Q}_0\right)^2}{g_n}\right\} \\
        &= P_0\left\{\frac{g_0-g_n}{g_n}\left(\frac{\bar{Q}_0}{g_0}-\frac{(\bar{Q}_0)^2}{g_0}-\frac{\bar{Q}_n(1-\bar{Q}_n)}{g_n}\right)+\frac{\left(\bar{Q}_n-\bar{Q}_0\right)^2}{g_n}\right\}\\
        & = P_0\left\{\frac{g_0-g_n}{g_n}\left(\frac{\bar{Q}_0}{g_0}-\frac{(\bar{Q}_0)^2}{g_0}-\frac{\bar{Q}_n(1-\bar{Q}_n)}{g_n} + \frac{\bar{Q}_0}{g_n}-\frac{\bar{Q}_0}{g_n} + \frac{\bar{Q}_0^2}{g_n}-\frac{(\bar{Q}_0)^2}{g_n}\right)+\frac{\left(\bar{Q}_n-\bar{Q}_0\right)^2}{g_n}\right\} \\
        &= P_0\left\{\frac{g_0-g_n}{g_n}\left(\left(\frac{\bar{Q}_0}{g_0}-\frac{\bar{Q}_0}{g_n}\right)-\frac{\bar{Q}_n}{g_n}+\frac{\bar{Q}_0}{g_n}\left(-\frac{(\bar{Q}_0)^2}{g_0}+\frac{(\bar{Q}_0)^2}{g_n}\right)+\left(\frac{(\bar{Q}_n)^2}{g_n}-\frac{(\bar{Q}_0)^2}{g_n}\right)\right)\right\}\\
        &\quad+P_0 \left\{\frac{\left(\bar{Q}_n-\bar{Q}_0\right)^2}{g_n}\right \}\\
        &= P_0\left\{\frac{g_0-g_n}{g_n}\left(\bar{Q}_0 \frac{g_n-g_0}{g_n g_0}+\frac{\bar{Q}_0-\bar{Q}_n}{g_n}+(\bar{Q}_0)^2 \frac{g_0-g_n}{g_n g_0}+\frac{(\bar{Q}_n-\bar{Q}_0)(\bar{Q}_n+\bar{Q}_0)}{g_n}\right)\right\}\\
          &\quad +P_0 \left\{\frac{\left(\bar{Q}_n-\bar{Q}_0\right)^2}{g_n}\right \}
    \end{split}
\end{equation}
As a result,

\begin{equation}
    \begin{split}
        R(P_n, P_0) &= P_0\left\{\frac{g_0-g_n}{g_n}\left(\bar{Q}_0 \frac{g_n-g_0}{g_n g_0}+\frac{\bar{Q}_0-\bar{Q}_n}{g_n}+(\bar{Q}_0)^2 \frac{g_0-g_n}{g_n g_0}+\frac{(\bar{Q}_n-\bar{Q}_0)(\bar{Q}_n+\bar{Q}_0)}{g_n}\right)\right\}\\
          &\quad +P_0 \left\{\frac{\left(\bar{Q}_n-\bar{Q}_0\right)^2}{g_n}\right \} + P_{0} (\frac{g_0-g_n}{g_n})\left(\frac{1-2 \bar{Q}_n}{g_n}+2\left(\bar{Q}_n-P_n\bar{Q}_n\right)\right)(\bar{Q}_0-\bar{Q}_n)\\
           &\quad -P_0\left(\bar{Q}_n-\bar{Q}_0\right)^2+\left(P_n \bar{Q}_n-P_0 \bar{Q}_0\right)^2
    \end{split}
\end{equation}

Only certain components of $R(P_n, P_0)$ can be bounded by a constant multiple of  $\| \bar{Q}_0-\bar{Q}_n\left\|_2\right\| g_0-g_n \|_2$. To ensure that $R(P_n, P_0)=o_P(n^{-1/2})$, we require that $g_n-g_0$,  $\bar{Q}_n-\bar{Q}_0$ and $P_n \bar{Q}_n-P_0 \bar{Q}_0$ approach $o_p(n^{-1/4})$ in $L_2(P)$ norm. This rate can be achieved using Highly Adaptive Lasso (HAL) with cross-validation, which has been shown to converge at a rate faster than $n^{-1/4}$ \cite{BenkeserHALE2016}. Unlike TMLE for common causal parameters, this estimator does not possess double robustness, as the structure of the remainder term does not allow for it.

\subsubsection{Second-order remainder of $\sigma$}
\begin{equation}
    \begin{split}
        \sigma(P) &= \mathbb{E}(D^{*}_{\psi_1, P} D^{*}_{\psi_0, P} \\
        &= \mathbb{E} \left(\frac{\mathbb{I}(A=1)}{g_P(1 \mid W)}\left(Y-\bar{Q}_P(A, W)\right)+\bar{Q}_P(1, W)-\psi_1(P) \right) \\ &\quad\left(\frac{\mathbb{I}(A=0)}{g_P(0 \mid W)}\left(Y-\bar{Q}_P(A, W)\right)+\bar{Q}_P(0, W)-\psi_0(P)\right)\\
        &= \mathbb{E} \left(\bar{Q}_P(1, W)-\psi_1(P)\right) \left(\bar{Q}_P(0, W)-\psi_0(P)\right) \\
        &=\mathbb{E}\bar{Q}_P(1, W)\bar{Q}_P(0, W)-\psi_0(P)\psi_1(P)
    \end{split}
\end{equation}

Let $\theta_1=\mathbb{E}\bar{Q}_P(1, W)\bar{Q}_P(0, W)$
$$
\sigma = \theta_1 - \mu_1\mu_0
$$
\begin{equation}
    \begin{split}
        D^{*}_{\sigma, P} &= D^*_{\theta_1, P} - \mu_0D^*_{\mu_1, P} - \mu_1D^*_{\mu_0, P}\\
        &=\bar{Q}^1\frac{1-A}{g_0}(Y-\bar{Q}^0) + \bar{Q}^0 \frac{A}{g_1}(Y-\bar{Q}^1) + \bar{Q}^1\bar{Q}^0-E(\bar{Q}^1\bar{Q}^0)- \mu_0\left( \frac{A}{g_1}(Y-\bar{Q}(A, W)) + \bar{Q}^1-\mu_1\right) \\
        &\quad- \mu_1\left( \frac{1-A}{g_0}(Y-\bar{Q}(A, W)) +\bar{Q}^0-\mu_0\right) \\
    \end{split}
\end{equation}
where
\begin{equation}
    \begin{aligned}
        D^*_{\theta_1, P} =\bar{Q}^1\frac{1-A}{g_0}(Y-\bar{Q}^0) + \bar{Q}^0 \frac{A}{g_1}(Y-\bar{Q}^1) + \bar{Q}^1\bar{Q}^0-E(\bar{Q}^1\bar{Q}^0)
    \end{aligned}
\end{equation}
Same conclusion is applied here like in Equation \ref{eq:remainder_linear_comb}. Since we already know the remainder terms for $\mu_1$ and $\mu_0$ is double robust, we only need to derive that of $\theta_1$ here. 

Therefore,
\begin{equation} \label{eq:remainder_theta}
    \begin{split}
        R_{\theta_1}(P_n, P_0) &= \theta_1(P_n)-\theta_1(P_0) + P_0D^*_{\theta_1, P_n}\\
        &=-P_0\bar{Q}^1_0\bar{Q}^0_0 + P_0\frac{1-A}{1-g_n}\bar{Q}^1_n(Y-\bar{Q}^0_n) + P_0\frac{A}{g_n}(Y-\bar{Q}^1_n)\bar{Q}^0_n  + P_0\bar{Q}^1_n\bar{Q}^0_n \\
        &=-P_0\bar{Q}^1_0\bar{Q}^0_0 + P_0\frac{g_0(0 \mid W)}{g_n(0\mid W)}Q^1_n(\bar{Q}_0^0-\bar{Q}^0_n) + P_0\frac{g_0}{g_n}\bar{Q}^0_n(\bar{Q}^1_0-\bar{Q}^1_n)  + P_0\bar{Q}^1_n\bar{Q}^0_n \\
        &= -P_0\bar{Q}^1_0\bar{Q}^0_0 + P_0\left(\frac{g_0(0 \mid W)}{g_n(0\mid W)}-1\right)\bar{Q}^1_n(\bar{Q}_0^0-\bar{Q}^0_n) + P_0Q^1_n(\bar{Q}_0^0-\bar{Q}^0_n) \\
       &\quad+ P_0\left(\frac{g_0}{g_n}-1\right)\bar{Q}^0_n(\bar{Q}^1_0-\bar{Q}^1_n)  + P_0\bar{Q}^0_n(\bar{Q}^1_0-\bar{Q}^1_n) + P_0\bar{Q}^1_n\bar{Q}^0_n \\
       &=P_0\left(\frac{g_0(0 \mid W)-g_n(0\mid W)}{g_n(0\mid W)}\right)Q^1_n(\bar{Q}_0^0-\bar{Q}^0_n) + P_0\left(\frac{g_0(1 \mid W)-g_n (1 \mid W)}{g_n}\right)\bar{Q}^0_n(\bar{Q}^1_0-\bar{Q}^1_n) \\
       &\quad- P_0\bar{Q}_0^1\bar{Q}_0^0 + P_0\bar{Q}^1_n\bar{Q}^0_0 + P_0\bar{Q}^1_0\bar{Q}^0_n-P_0\bar{Q}^1_n\bar{Q}^0_n\\
       &=P_0\left(\frac{g_0(0 \mid W)-g_n(0\mid W)}{g_n(0\mid W)}\right)\bar{Q}^1_n(\bar{Q}_0^0-\bar{Q}^0_n) + P_0\left(\frac{g_0(1 \mid W)-g_n (1 \mid W)}{g_n}\right)\bar{Q}^0_n(\bar{Q}^1_0-\bar{Q}^1_n) \\
       &\quad -P_0 (\bar{Q}^1_0 - \bar{Q}^1_n)(\bar{Q}^0_n-\bar{Q}^0_0)\\
    \end{split}
\end{equation}

The third term in the Equation \ref{eq:remainder_theta} does not exhibit the cross‑product structure that yields double robustness. As a result, achieving asymptotic efficiency hinges on correctly specifying (and consistently estimating) the outcome‑regression model; any misspecification will prevent the remainder from shrinking at the required $o_P(n^{-1/2})$ rate.

With these remainder terms appropriately controlled, we conclude that $R(P_n, P_0)$ is $o_P(n^{-1/2})$. Therefore, $\Sigma^{2. TMLE}_n$ is an asymptotically efficient estimator of $\Sigma^2_n$ under the stated conditions.



\section{Distribution of variance estimates under simple simulation}  \label{appendix:dist_var}

\begin{figure}[H]
    \centering
    \includegraphics[width=\linewidth]{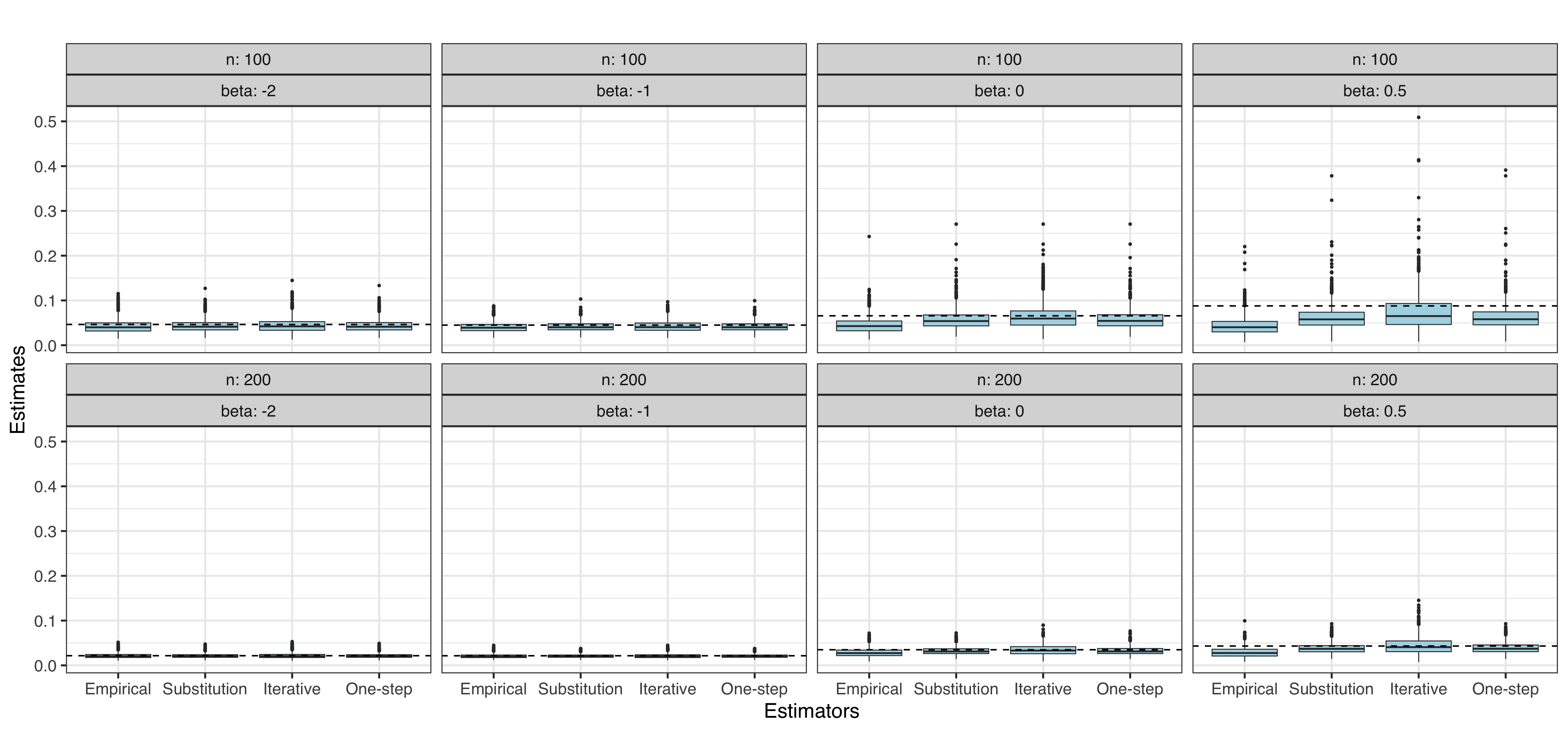}
    \caption{Distribution of variance estimates of different variance estimators compared to the Monte-Carlo variance under small sample sizes.}
    \label{figure_A4_1}
\end{figure}

\begin{figure}[H]
    \centering
    \includegraphics[width=\linewidth]{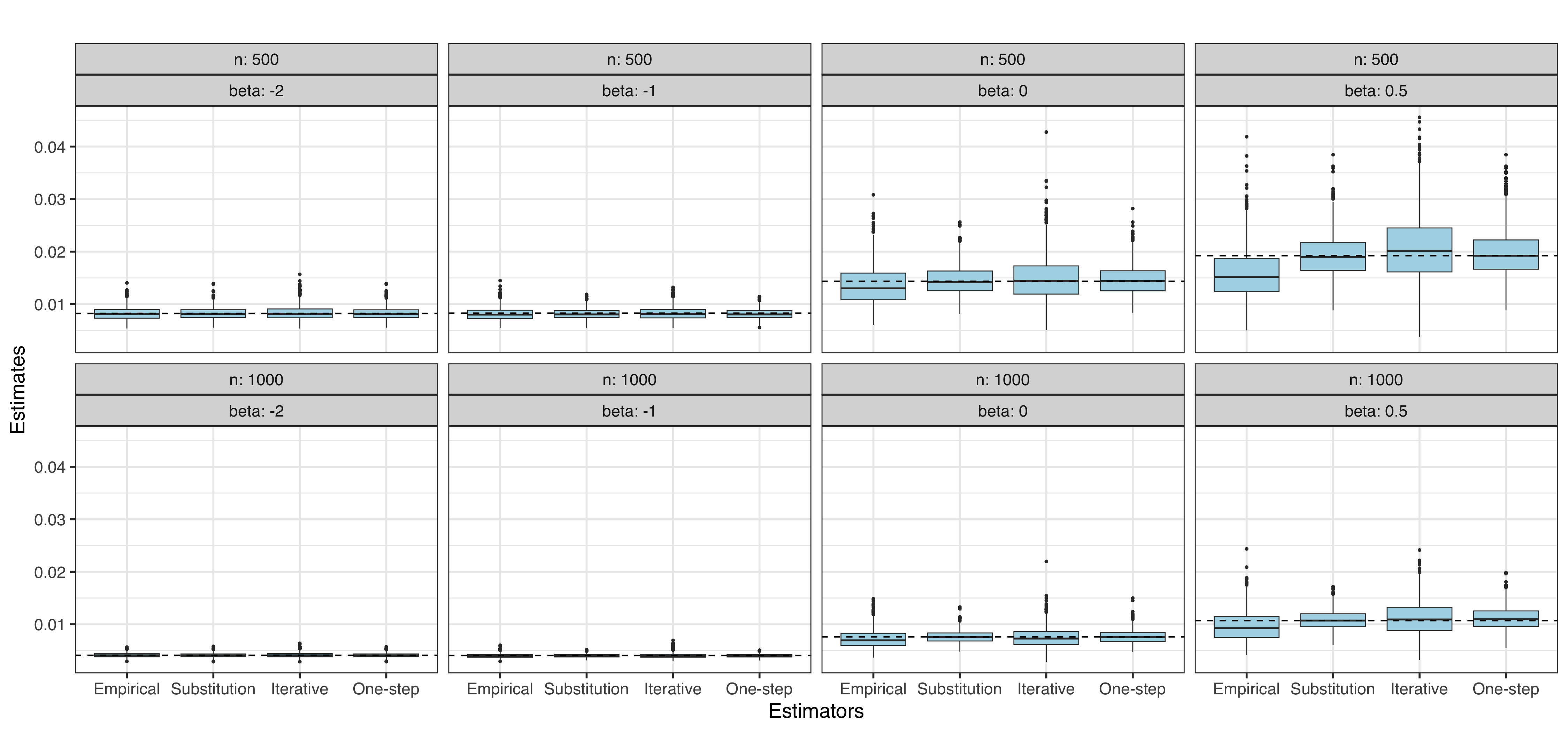}
    \caption{Distribution of variance estimates of different variance estimators compared to the Monte-Carlo variance under large sample sizes}
    \label{figure_A4_1}
\end{figure}
\section{Simulation results for other DGDs} \label{appendix:others}

\subsection{DGD with misspecified outcome regression model}
We also conduct simulations under the condition where we misspecify the outcome regression model, which is the Q form parameter in tmle package when we get the initial estimates of $\bar{Q}_0$. Other settings remain unchanged to the original simulation. 
\subsubsection{Results on the performance of different estimator}
\begin{figure}[H]
    \centering
    \includegraphics[width=\linewidth]{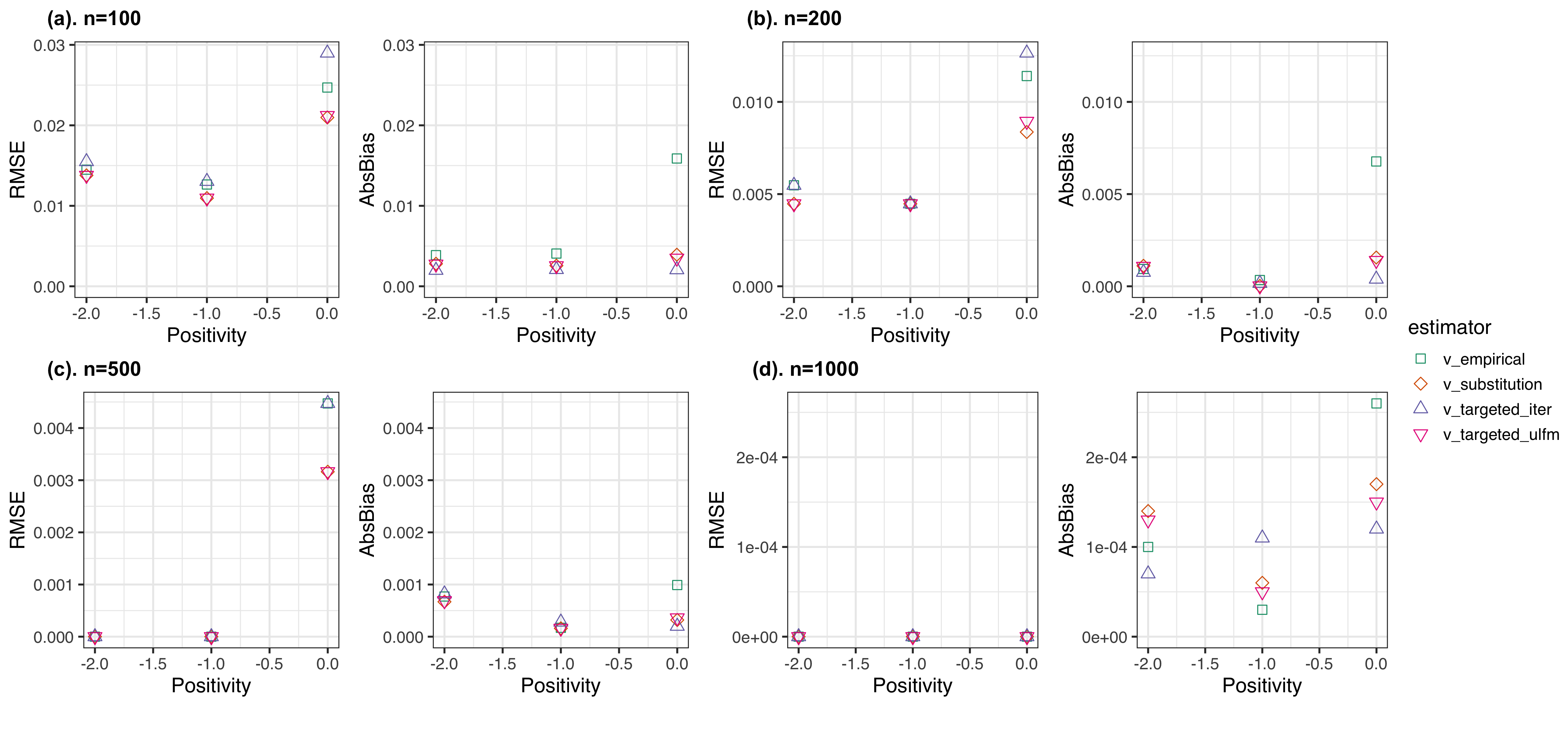}
    \caption{RMSE and absolute bias of each variance estimator compared to $\Sigma^2(P_0)$ under various sample sizes and positivity (Misspecified Q)}
    \label{fig:enter-label}
\end{figure}

\subsubsection{Results on the inference of log(CRR)}
\begin{figure}[H]
    \centering
    \includegraphics[width = \linewidth]{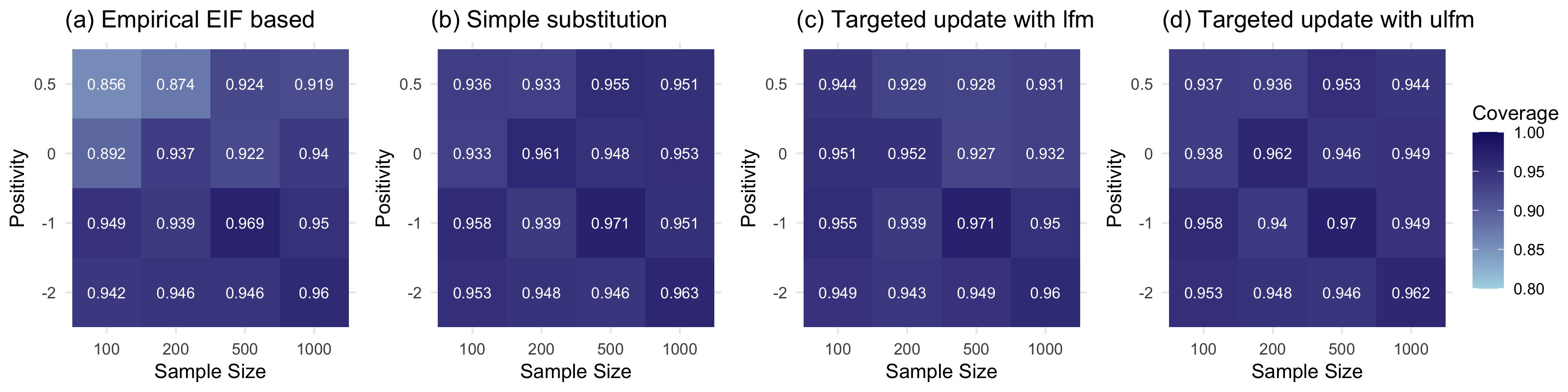}
    \
    \caption{Coverage rates for each variance estimator for the TMLE estimator of log(CRR) under various sample sizes and positivity ($\beta_p$) (Misspecified Q)}
    \label{figure_A4_3}
\end{figure}

\begin{figure}[H]
    \centering
    \includegraphics[width=\linewidth]{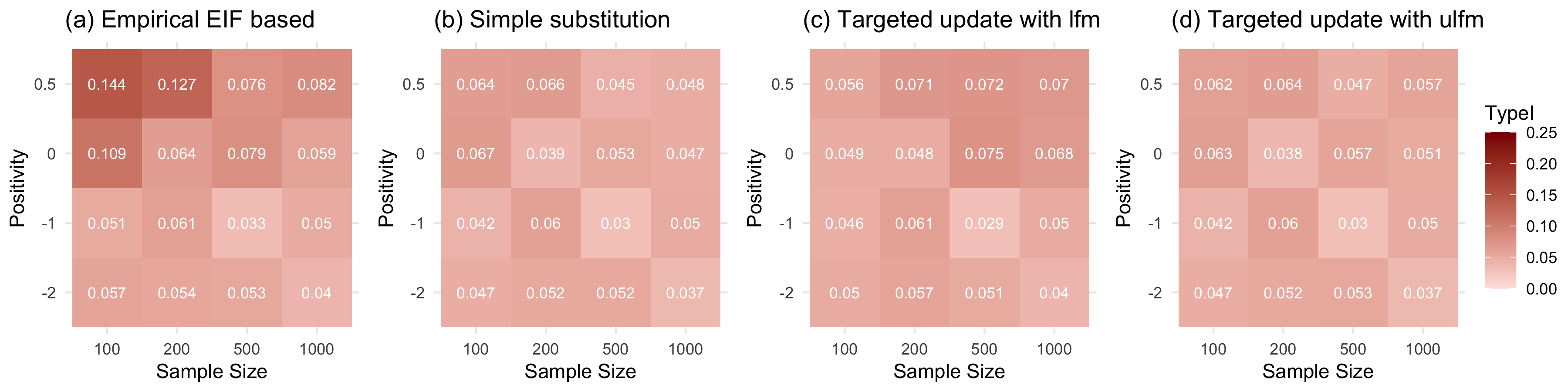}
    \caption{Type I error rates for each variance estimator for the TMLE estimator of log(CRR) under various sample sizes and positivity ($\beta_p$) when the null is true: $\Psi(P_0)=0$ (Misspecified Q)}
    \label{fig:enter-label}
\end{figure}

\subsection{Complex setting} 
The simulation with more complex relationships different from the original one in the propensity score and conditional outcome formula by including interaction terms between covariates, as shown below:
\begin{align*}
W_{1} & \sim U(0, 1), \\
W_{2} & \sim U(0, 1), \\
W_{3} & \sim U(0, 1), \\
A &\sim Bernoulli(p), \text{where } p =  \mathrm{logit}^{-1}\left(\beta_p - (\beta_p + 2.5)W_{1} + 1.75W_{2} + (\beta_p + 3.2)W_{3} - 0.75W_1W_2 + 0.75W_2^2\right)\\
Y & \sim Bernoulli(q), \text{where } q = \mathrm{logit}^{-1}\left(0.1 + 0.1W_{1} + 0.1W_{2} + 0.2W_{3} -0.5 W_1W_3 + 0.3W_1^2+\beta_{\psi_0}A\right)
\end{align*}

During the implementation of methods, we use the default SuperLearner libraries in the tmle packages to get the initial estimates of $\bar{Q}_0$ and $g_0$, which are the generalized linear model, Bayesian additive regression trees and Generalized Linear Model via Elastic Net for $Q$ and generalized linear model, Bayesian additive regression trees and generalized additive model for $g$.

\subsubsection{Results on the performance of different estimators}
\begin{figure}[H]
    \centering
    \includegraphics[width=\linewidth]{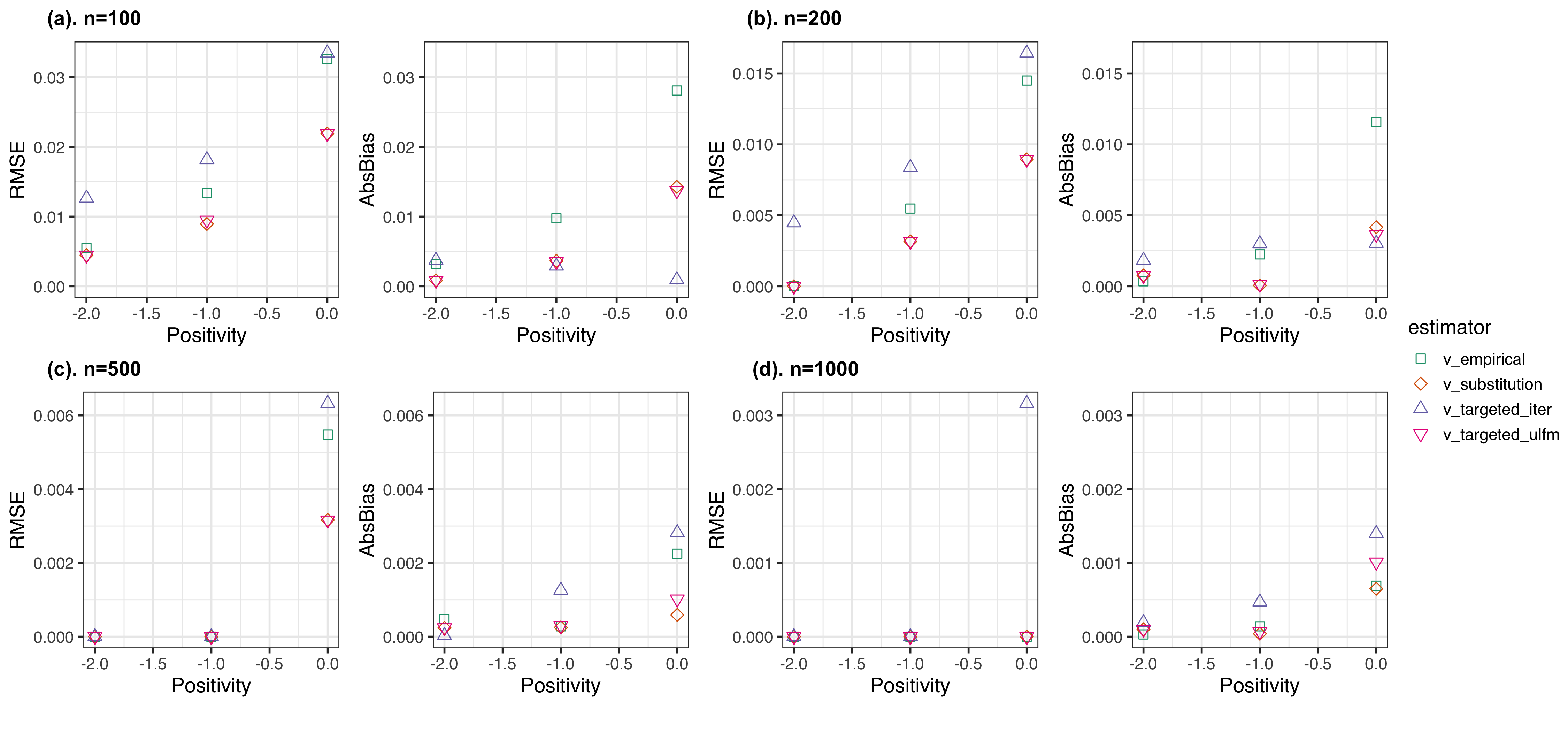}
    \caption{RMSE and absolute bias of each variance estimator compared to $\Sigma^2(P_0)$ under various sample sizes and positivity (complex simulation).}
    \label{fig:enter-label}
\end{figure}

\begin{figure}[H]
    \centering
    \includegraphics[width=\linewidth]{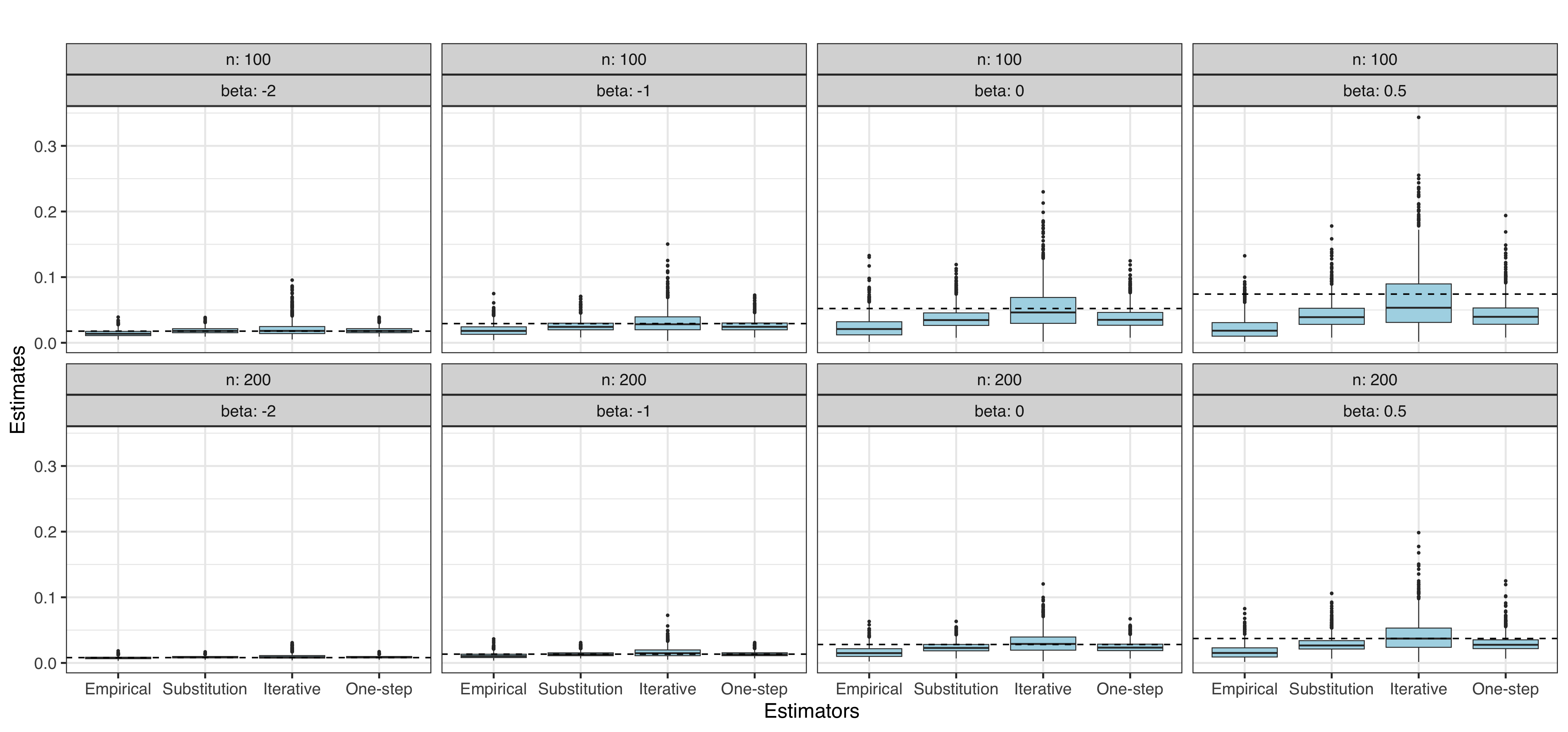}
    \caption{Distribution of variance estimates of different variance estimators compared to the Monte-Carlo variance under small sample sizes (complex simulation).}
    \label{fig:enter-label}
\end{figure}

\begin{figure}[H]
    \centering
    \includegraphics[width=\linewidth]{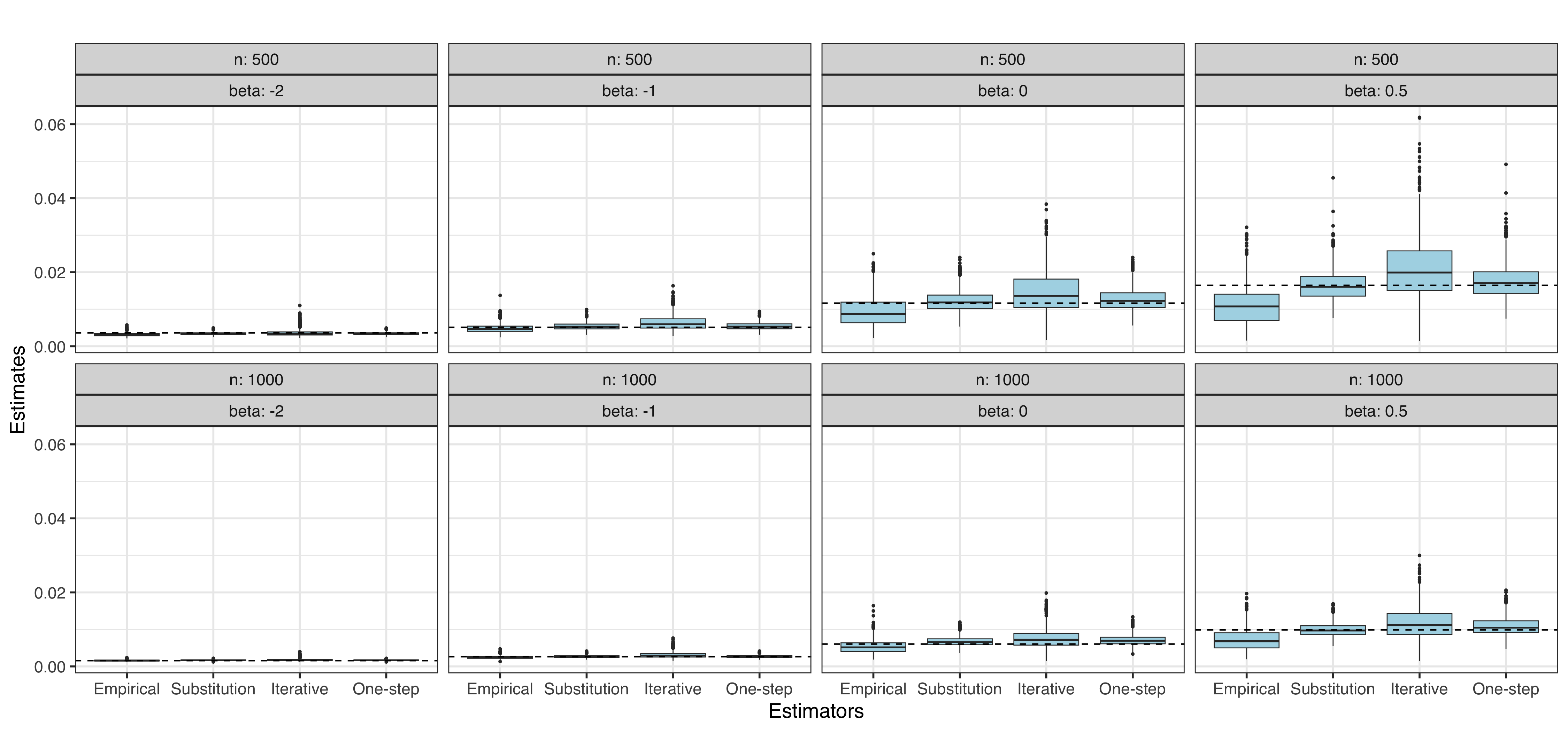}
    \caption{Distribution of variance estimates of different variance estimators compared to the Monte-Carlo variance under large sample sizes (complex simulation).}
    \label{fig:enter-label}
\end{figure}

\subsubsection{Results on the inference of log(CRR)}

\begin{figure}[H]
    \centering
    \includegraphics[width = \linewidth]{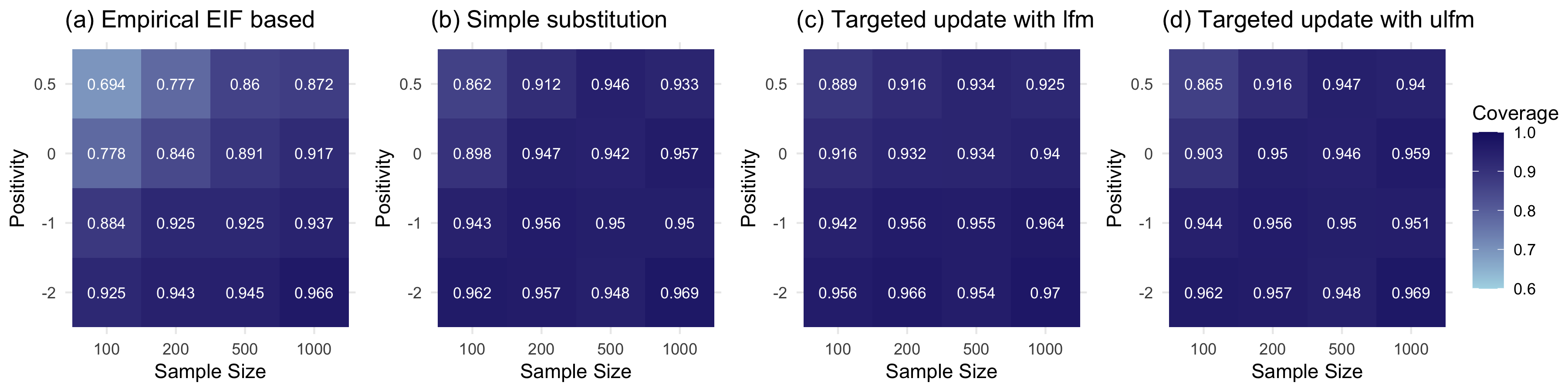}
    \caption{Coverage rates for each variance estimator for the TMLE estimator of log(CRR) under various sample sizes and positivity ($\beta_p$) (complex simulation).}
    \label{figure_A4_1}
\end{figure}

\section{Levels of propensity truncation for $\beta_p$}

Table \ref{tab:prop_truncation} shows the proportion of observations with estimated propensity scores ($g_0$) truncated at the level of 0.025.
\begin{table}[h]
\centering
\begin{tabular}{ccccc}
\toprule
\textbf{Simulation}&   -2&-1&0& 0.5\\
\midrule
Simple&   0&0&0.024& 0.065\\
Complex&   0&0&0.042& 0.131\\
\end{tabular}
\caption{Proportion of observations with truncated $g_0$ at each $\beta_p$ under the simple and complex data-generating distributions.. }
\label{tab:prop_truncation}
\end{table}

\end{document}